\documentclass[a4paper,10pt,aps]{revtex4-1}

\usepackage{amsmath}
\usepackage{bbm}
\usepackage{graphicx,amssymb,bm,latexsym,color,epsf,subfig}
\usepackage{breqn}
\usepackage[colorlinks]{hyperref} 

\newcommand{\be}{\begin{equation}}
\newcommand{\ee}{\end{equation}}
\newcommand{\bea}{\begin{eqnarray}}
\newcommand{\eea}{\end{eqnarray}}

\makeatletter
\let\cat@comma@active\@empty
\makeatother

\begin{document}


\title{Higher-derivative four-dimensional sine-Gordon model}


\author{Matteo F. Bontorno}
\affiliation{Dipartimento di Fisica e Astronomia “Ettore Majorana”, Universit\`a di Catania, 64, Via S. Sofia, I-95123 Catania, Italy;}
\affiliation{Scuola Superiore di Catania, Universit\`a di Catania, 9, Via Valdisavoia, I-95123 Catania, Italy;}
\affiliation{Centro Siciliano di Fisica Nucleare e Struttura della Materia, Catania, Italy.}

\author{G.G.N. Angilella}
\affiliation{Dipartimento di Fisica e Astronomia “Ettore Majorana”, Universit\`a di Catania, 64, Via S. Sofia, I-95123 Catania, Italy;}
\affiliation{Scuola Superiore di Catania, Universit\`a di Catania, 9, Via Valdisavoia, I-95123 Catania, Italy;}
\affiliation{INFN, Sezione di Catania, Via Santa Sofia 64, 95123 Catania, Italy.}
\affiliation{Centro Siciliano di Fisica Nucleare e Struttura della Materia, Catania, Italy.}

\author{Dario Zappal\`a}
\affiliation{INFN, Sezione di Catania, Via Santa Sofia 64, 95123 Catania, Italy;}
\affiliation{Centro Siciliano di Fisica Nucleare e Struttura della Materia, Catania, Italy.}

\begin{abstract}
\vskip 30pt
\centerline{ABSTRACT}
\vskip 10pt
The phase structure of a higher derivative sine-Gordon model in four dimensions is analysed. It is shown that 
the inclusion of a relevant  two-derivative term in the action significantly modifies some of  the results obtained by neglecting this 
operator, and the final picture  is substantially different from the one describing the phase diagram associated with 
the two-dimensional Berezinskii-Kosterlitz-Thouless  (BKT) transition. The study is carried out with the help 
 of the Renormalization Group (RG) flow equations, determined  for a set of  three parameters, and numerically 
 solved both for  a truncated series expansion approximation, and for the complete set of equations.
 In  both cases, a continuous line of fixed points,  terminating at a particular point presenting  universal properties,
 is found, together with a manifold that separates two phases, roughly characterized by the sign of the 
 coupling $\widetilde z_k$ associated with this newly included operator.  While the phase corresponding to
 $\widetilde z_k>0$ shows some pathologies, the one with $\widetilde z_k<0$ has a well-behaved infrared limit,
where the  system reduces to a Gaussian-like  model.  
We also briefly comment about the possibility that  our model could capture some of the qualitative features of the ultraviolet (UV) 
critical manifold of conformally reduced gravity.
\end{abstract} 


\maketitle

\setcounter{page}{2}

\section{introduction}

Field  theories containing  higher derivative kinetic terms  have received long-standing attention, as they present  the  
remarkable advantage of an improved ultraviolet (UV) regime  due to the smoothening of loop integrals in the high loop 
momentum limit,  because of the higher powers of momentum  in the propagators. Therefore, 
various applications have been considered so far to  theories that are otherwise non-renormalizable, 
such as the  case of the field theoretical formulation of gravity
 \cite{thirring,Pais:1950za,Stelle:1976gc,deUrries:1998obu,Hawking:2001yt,Collins1}.

Actually, this approach suffers from a fundamental inconvenience, 
as theories  that contain more than two time derivatives  are affected  by the Ostrogradski instability, 
associated with the  violation of unitarity \cite{deUrries:1998obu,Woodard:2006nt}.
This means that higher derivative theories containing the same number of space and time derivatives, a property 
that is essential to comply with Lorentz symmetry, unavoidably must bear with the violation of other basic principles
which are characteristic of a fundamental theory.

In this sense, field theories with an anisotropic structure, i.e. with actions containing two derivatives with respect to the time coordinate,
 but  $2\tau$ derivatives with respect to the space coordinates ($\tau$ is usually indicated  as the anisotropy exponent), have been analysed. 
So, for instance, the  Ho\v{r}ava-Lifshitz gravity \cite{horava,Horava2009}  is  a renormalizable anisotropic gravitational theory, 
constructed on the basis of these motivations (for a more recent version of this formulation see Refs.~\cite{Barv1,Barv2,Barv3,Barv4}).
In addition, anisotropic field  theories on flat spacetime were  also formulated on the same grounds 
\cite{Anselmi:2007ri,Iengo,Dhar:2009dx,horava:ym,eune,alexandre,Kikuchi,Chao:2009tqf,Solomon:2017nlh,zap1,zap2}.
Clearly, this class of actions does not fulfill Lorentz symmetry, that can eventually be recovered at low energy as an emergent property, 
but it is essential to require that the related theoretical predictions are not in contradiction with  
the observational constraints \cite{chenhuang, Ellis:2018lca}. 

However, the study of isotropic higher derivative theories, i.e. with actions containing the same number of space and time derivatives,
has nevertheless been pursued, because they can still be regarded as effective field theories, possibly generated by some integration 
of more fundamental degrees of freedom and, therefore, not necessarily constrained by all fundamental principles,
that are still  helpful as examples of renormalizable  theories.   In addition, the relevance of these isotropic, 
higher derivative theories also comes from the fact that, when Wick-rotated, they acquire the usual physical interpretation  
of a statistical model  in a $d$-dimensional space.

The renormalization of these higher-derivative theories is regulated by the presence of a particular class of fixed points, called Lifshitz points, 
that may show up in both isotropic and anisotropic actions  \cite{Horn}. Concerning statistical systems, they are 
typically associated to the existence of tricritical points in the phase diagram, which signal the coexistence of three phases, including one  
that presents a non-uniform, modulated ground state.
This has several realisations in condensed matter, such as magnetic systems, but also polymer mixtures, liquid crystals, high-Tc superconductors 
(for reviews on this subject see \cite{erzan,sak,hornrev,Diehl3}), as well as in the analysis of dense quark matter with the realization of 
unconventional phases \cite{casal2,nardulli,buballa,pisarski,Pisarski:2020dnx}.

From the point of view of renormalization,  the appearance of  Lifshitz points  comes from a change of the scaling dimensions
of the various couplings  in the action with respect to the canonical dimensions,  due to the presence of a higher derivative term
which set the new scaling dimension of the field,  different from its canonical dimension determined instead by the two derivative kinetic term.
The change of scaling dimensions modifies the nature of the Renormalization Group (RG) flow equations and the structure of fixed points 
that regulate the renormalization properties of the theory  \cite{Horn}.

In the past years, this modification of the scaling dimension was exploited in a series of 
papers \cite{Bonanno:2014yia,Zappala:2017vjf,Zappala:2018khg}, 
to establish a relation between the four-dimensional (4D) isotropic  higher-derivative scalar theories, 
characterized by the kinetic term $W\,\partial^2 \phi\partial^2 \phi$, and the two-dimensional (2D) standard theory 
with kinetic term  $Z\,\partial \phi\partial \phi$, as both models, in the presence of a continuous 
 internal symmetry $O(N)$,  show typical features of systems defined in a $\overline{d}$-dimensional space,
$\overline{ d}$  being equal to their lower critical dimension. 

The  non-perturbative approach used in this investigation is the RG analysis, based on the Functional RG 
flow equations \cite{Wetterich:1992yh, Morris:1994ie, Berges:2000ew},  which, for instance, was extensively used to 
produce a  comprehensive picture of the universal properties of $O(N)$ field theories as a function of the dimension 
$d$ and the index $N$ \cite{Codello2015,Defenu2015,Defenu:2017el},
as well as of topological phase transitions such as the 2D Berezinskii - Kosterlitz -Thouless (BKT)  transition  
\cite{grater,jame,Jakubczyk2014,defenu2017-1,krieg} 
and of the sine-Gordon model \cite{Nandori2001,Nagy2009,Nandori09,Daviet:2018lfy,Jentsch:2021trr,Nandorifin}, 
that belongs to the same universality class.

Then, in \cite{zapp,defenu}, the RG technique developed for the 2D topological transitions,
 was exported to 4D systems, by first mapping an $O(2)$-symmetric  four-derivative model onto a higher 
 derivative sine-Gordon model (analogous  to the well established mapping between the XY  and  the  
 sine-Gordon model in 2D \cite{minnhagen}), and then studying the RG flow equations obtained  for the latter.
 The picture emerging   from this analysis  remarkably shows several features of the  BKT universality 
 class \cite{berezinskii71,kosterlitz72,kosterlitz73,Jose1977},
such as the presence of a line of fixed points terminating at an end point associated with a  
universal value  of the  anomalous dimension  $\eta$.

In this paper, we plan to further  investigate this issue, as in \cite{defenu} it was pointed out that, 
when passing from a 2D  to a 4D system, a few relevant differences appear, besides the mentioned similarities.
In fact, the 4D sine-Gordon model RG flow equations present a sign flip with respect to the 2D case, that unavoidably modifies  the phase diagram.
Here, we reconsider the analysis of \cite{defenu} and, besides deriving the flow equation within a different approach to provide an additional check of 
the change of sign in the RG equations, we enlarge our model, by including
 an additional  two-derivative kinetic term into the 4D sine-Gordon model, which was previously neglected,
 in order to get a more comprehensive description of such a universality class.
Actually, the inclusion of such a term is a crucial point, as it turns out to be a relevant operator, 
according to the computation of its scaling dimension and, consequently,
it plays an essential role in determining the actual infrared (IR) limit of the RG  flow and the complete structure of the phase diagram.

In Sec.\,\ref{themodel}, we introduce the specific model and the associated complete RG flow equations. We also summarise the results obtained 
in the approximation  studied in \cite{defenu}, to better clarify the improvements achieved in this paper.
Sec.\,\ref{expanded} is devoted to the results of a truncated version of the RG flow equations
(after an expansion in powers of the couplings around the line of fixed points), 
which is simpler to handle but sufficient to grasp the main features of our model.
In Sec.\,\ref{complete} the results derived from the full RG equation are investigated.
Then, in Sec.\,\ref{conclusion} we discuss the possible future perspectives of the present analysis.
Finally, Appendix \ref{appendixa} contains the main details of the derivation of the RG flow equations,
 while in Appendix \ref{appendixb}
an alternative derivation of the RG equations, based on a perturbative approach, is presented.

\section{Generalized sine-Gordon Model }
\label{themodel}

In the following we shall discuss the phase structure of the Euclidean 4D model defined by
 ($\Delta\equiv \partial_\mu\partial_\mu$)
\begin{equation}
S [\Phi] = \int d^4 {x} \; \Big [ \frac{ w }{2} \, \Delta \Phi ({x})
  \Delta \Phi ({x}) + \frac{Z}{2} \, \partial_\mu \Phi ({x})  \partial_\mu \Phi ({x})  + g\, (1- \cos \beta \, \Phi )   \Big ]
\label{model}
\end{equation}
which generalizes the sine-Gordon model discussed in \cite{defenu}, where the  two-derivative term is absent by imposing $Z=0$. 
In Eq. (\ref{model}), the parameter $\beta$ is included to keep the product $ \beta \, \Phi$  adimensional, 
whatever the engineering dimension of the field $\Phi$. Clearly one can get rid of $\beta$, by replacing in  Eq. (\ref{model}) : 
\begin{equation}
u= w / \beta^2   \;\;\;\; ; \;\;\;\;\;\;\;\;   z= Z / \beta^2   \;\;\;\; ; \;\;\;\;\;\;\;\;   \varphi = \beta\, \Phi  \; . 
\label{redefine}
\end{equation}
Then, the actual value of $\beta$ can be absorbed in the scale of the Lagrangian, which forms the argument of the integral defining the action in Eq. \eqref{model}. 
A similar situation occurs when considering a simple model for the degenerate electron gas, including the electronic kinetic energy and Coulomb repulsion, 
as well as a uniform positive background, where all lengths and wavevectors are scaled using the Wigner radius $r_s$. In that case, the overall Hamiltonian 
acquires a factor of $1/r_s^2$, while the (renormalized) interaction term exhibits a relative scale of $r_s$ with respect to the kinetic energy term
(see e.g. Eq. (3.24) of \cite{fetter}).

However, for the specific case in Eq. \eqref{model},  $\beta$ is to be regarded as a fixed external parameter that does not get modified by the RG transformations 
along the flow, as stated in \cite{Nandorifin}. This, in turn, means that, once the  redefinitions (\ref{redefine}) are inserted in  Eq. (\ref{model}), 
any other rescaling of the field does 
imply a genuine physical scale change in the model. In fact, in the usual 2D sine-Gordon model, different  values of the field renormalization 
(i.e. the coefficient of the term containing two derivatives of the field), produce  physically distinct fixed points of the system unlike, for instance,  
the critical exponents of the 3D Wilson-Fisher fixed point that are not modified by a global rescaling of the field renormalization, which is
compensated by an unobservable redefinition  of the field, as shown, e.g., in \cite{bonannozapp}.

The model with $Z=0$ analysed in \cite{defenu}, shows many features that are characteristic of the 2D  BKT transition,
namely the presence of a line of fixed points parametrized by a variable which undergoes a  universal jump from a finite value to zero, 
at a specific critical point. Unfortunately, other details, and in particular the full phase diagram, of the BKT transition were not recovered in 
\cite{defenu}, because of the differences in the flow equations obtained in that particular case with respect to those governing the 2D BKT transition.
Hence, after deriving the flow equations for the model (\ref{model}) we shall reconsider the case with $Z=0$, analyzing in detail its flow.

The  study of the phase structure of our model  can be performed by means of  the  Functional Renormalization Group (RG),
where  the effect of  fluctuations is represented in terms of a functional differential flow equation which describes the evolution with a momentum scale $k$
of the various parameters of the average effective action functional, defined in accordance with Eqs. \eqref{model} and \eqref{redefine}
\begin{equation}
\Gamma_k[\varphi] = \int d^{\rm d} {x} \; \Big [
\frac{u_k}{2}\partial^2\varphi\partial^2\varphi+\frac{{z}_k}{2}\partial_{\mu}\varphi\partial_{\mu}\varphi+{g}_k(1-\cos{\varphi}) \Big ]
\label{gammak}
\end{equation}
where, without loss of generality, we eliminated the parameter $\beta$ by making use of the replacements displayed in Eq. \eqref{redefine} 
(but  we shall come back to the original form in  Eq. \eqref{model} at the end of Sec.\,\ref{complete}  to discuss the IR limit of the model). 
We also generalised  Eq.  \eqref{gammak} to d dimensions.
This is done for computational convenience, as discussed in Appendix 
\ref{appendixa}, and the limit $d\to 4^+$ is taken only as the last step of our calculation. 
Then,  the parameters, $u_k$, $z_k$, $g_k$ in \eqref{gammak} show a new dependence on the energy scale $k$, as we let them  evolve
 according to  a set of RG differential equations, starting from a set of values assigned to the 
microscopic model at the UV scale $k=\Lambda$, down to the  IR region for  $k\to 0$.  In particular, we shall resort to the  RG flow equation 
\cite{Berges:2000ew}
\begin{equation}\label{wettflow}
\partial_t\Gamma_k[\varphi]=\frac{1}{2} \; {\rm Tr} \frac{\partial_t\, R_k}{\Gamma_k^{(2)}[\varphi] + R_k}
\end{equation}
with  $t= \log ( \Lambda/k)$,  and the trace is taken over spatial and momentum degrees of freedom;
 $\Gamma_k^{(2)}[\varphi] $ is the second functional derivative of $\Gamma_k[\varphi]$ with respect to the field $\varphi$; 
 finally $R_k $ is the scale dependent regulator function which effectively generates the progressive integration of the UV modes along the  flow.
 For our purposes, it  is sufficient to  select the simple form \cite{defenu} 
\begin{equation}\label{Rk}
R_k = k^4 \;.
\end{equation}

Then, one has to suitably project  the full flow equation (\ref{wettflow}), in order to determine the flow of the three 
parameters $u_k,\, z_k,\,g_k$. In particular, the projection on $x$-independent field configurations selects the flow 
equation  for the potential  $V_k(\varphi)= \int d^4 {x} \; g_k \, (1- \cos \varphi )$
\begin{equation}\label{Vflow}
\partial_t V_k(\varphi) =\frac{1}{2}\int\frac{d^d q}{(2\pi)^d} \;   (\partial_t R_k) \; G_k (q) \; ,
\end{equation}
where 
\begin{equation}\label{propag0}
G_k (q) =  \Big ( \Gamma_k^{(2)}+R_k \Big )^{-1}  \; ,
\end{equation}
and $V_k''(\varphi)$ is the second derivative of the potential with respect to $\varphi$.
In order to determine the flow of $ g_k$, it is sufficient to select  the coefficient of $\cos \varphi$, which is 
realised by applying  
on both sides of Eq. (\ref{Vflow})  the projector \cite{defenu} :
\begin{equation}
P_1=-\frac{1}{\pi}\int_{-\pi}^{\pi}d\varphi\cos{\varphi} \; .
\label{proj1}
\end{equation}

On the other hand, the determination of the flow equations for the coefficients of the derivatives of the field, namely $u_k$ and $z_k$, 
requires one more step.
In fact, one has first to determine the flow of the two-point function   $\Gamma_k^{(2)}[\varphi]$ by differenziating  Eq. (\ref{wettflow}) twice with 
respect to the field $\varphi$:
\begin{eqnarray}
 &&\partial_t\Gamma_k^{(2)}(-p,p,\varphi)=
 \nonumber\\ 
 &&\int \frac{d^dq}{(2\pi)^d}\; \frac{  (\partial_t R_k  )  }{2}   \left[2\frac{\Gamma_k^{(3)}(p,q,p-q,\varphi)}{\Gamma_k^{(2)}(-q,q,\varphi)+R_k} G_k(p+q)
\frac{\Gamma_k^{(3)}(p,p+q,-q,\varphi)}{\Gamma_k^{(2)}(-q,q,\varphi)+R_k}-\frac{\Gamma_k^{(4)}(p,q,-q,-p)}{(\Gamma_k^{(2)}(-q,q,\varphi)+R_k)^2} \right] \; ,
\label{gamma2flow}
\end{eqnarray}
where we explicitly displayed  the dependence on the specific momenta, of the two-, three-, four-field derivatives of the average action: 
$\Gamma_k^{(2)} , \, \Gamma_k^{(3)} , \, \Gamma_k^{(4)}$. However, according to the ansatze in Eq. \eqref{gammak}, 
in the following we shall compute the $n$-point functions in Eq. \eqref{gamma2flow},
as the $n$-derivative of the action in \eqref{gammak}. In this approximation, $\Gamma_k^{(3)}$ and  $\Gamma_k^{(4)}$
are identified with the third and fourth derivatives of the potential $V_k(\varphi)$, respectively, and do not carry any dependence on 
the external momentum $p$ of Eq. \eqref{gamma2flow}, while,  according to the definition in  Eq. \eqref{propag0},
for the two-point function we have
\begin{equation}\label{propag}
G_k (q) =  \frac{1}{u_k \, q^4 + z_k \, q^2 + V_k''(\varphi)  + R_k} \;.
\end{equation}

The flow equations for  $u_k$ (or $z_k$) are obtained by taking four (two) derivatives 
 with respect to the external momentum $p$ and then  applying  on both sides of Eq. (\ref{gamma2flow})  the projector 
 \begin{equation}
 P_0=\frac{1}{2\pi}\; \int_0^{2\pi}d\varphi
\label{proj0}
\end{equation} 
which yields 
\begin{equation}\label{uk}
    \partial_tu_k=\frac{P_ 0}{4!}\lim_{p\to 0}\int\frac{d^dq}{(2\pi)^d}\partial_tR_kG \
(q)^2V_k^{(3)}(\varphi)^2\frac{d^4}{dp^4}G(p+q)
\end{equation}
\begin{equation}\label{zk}
\partial_tz_k=\frac{P_ 0}{2!}\lim_{p\to 0}\int\frac{d^dq}{(2\pi)^d}\partial_tR_kG(q)^2V_k^{(3)}(\varphi)^2\frac{d^2}{dp^2}G(p+q) \;.
\end{equation}
Note that the last term on the right hand side of Eq.  \eqref{gamma2flow}, proportional to $\Gamma_k^{(4)}$, does not contribute to Eqs. \eqref{uk},
\eqref{zk},  because in our scheme it has no dependence on the external momentum  $p$.

Of course, the significant RG flow is obtained for the variables suitably expressed in units of the running scale $k$. 
In our case, we expect the non-perturbative dynamics to be driven by the higher derivative term in Eq. \eqref{gammak}. 
This, in turn, means
that the coefficient of the higher derivative term, $u_k$ has zero scaling dimension, and consequently also the field $\varphi$ has zero scaling dimension.
Then, after a  simple dimensional analysis, we define the new (scaling) dimensionless variables
\begin{equation}\label{tilde}
\widetilde z_k \equiv \frac{z_k}{k^2}  \;\;\;\; ; \;\;\;\; \widetilde g_k \equiv \frac{g_k}{k^4} 
\end{equation}
while, since $\widetilde u_k  =u_k$, in this case we do not change notation and keep the dimensionless variable $u_k$.
This redefinition of the scaling dimensions, different from the usual case where the coefficient of the two-derivative term $z_k$ is 
taken as dimensionless, leads to the presence of Lifshitz points, and a peculiar UV structure even for the  simplified version of model \eqref{model},
where the two derivative term is omitted \cite{defenu}, and here we analyse the case where $z_k$ is turned on.

The  RG calculation yields the three beta-functions 
\begin{eqnarray}\label{rgsys}
&&\partial_t \, u_k =\beta_u (u_k, \widetilde z_k, \widetilde g_k) \nonumber\\
&&\partial_t \, \widetilde z_k =\beta_z (u_k, \widetilde z_k, \widetilde g_k) \nonumber\\
&&\partial_t \, \widetilde g_k =\beta_g (u_k, \widetilde z_k, \widetilde g_k)  \; .
\end{eqnarray}
Details of the computation are provided  in Appendix \ref{appendixa}  and the explicit form of 
$\beta_u (u_k, \widetilde z_k, \widetilde g_k)$,  $\beta_z (u_k, \widetilde z_k, \widetilde g_k)$ and $\beta_g (u_k, \widetilde z_k, \widetilde g_k)$
is given respectively in Eqs. \eqref{flowu}, \eqref{flowz},  \eqref{flowg}.

Before considering the complete problem of three coupled equations, let us summarize the main features of the case where the two derivative term is omitted,
i.e. $z_k=0$ is kept fixed while focussing on the flow of  $u_k$ and $\widetilde g_k$ only. If one sets $z_k=0$ in  \eqref{flowu},   \eqref{flowg}, the 
corresponding equations are  significantly simplified and they coincide with those derived in \cite{defenu}:

\begin{equation}
\label{flowuredu}
 \partial_tu_k=- \frac{\widetilde{g}^2_k}{160\pi^2\sqrt{(1-\widetilde{g}^2_k)^3}}
\end{equation}
\begin{equation}
\label{flowgredu}
\partial_t\widetilde{g}_k=4\widetilde{g}_k-\frac{1}{8\pi^2u_k\widetilde{g}_k}\left(1-\sqrt{1-\widetilde{g}_k^2}\right).
\end{equation}
It should be noted however that the constraint $\widetilde z_k=0$  does not imply $\partial_t \widetilde z_k=0$, and we shall come back to this point in Sec. \ref{expanded}.

Eqs. \eqref{flowuredu} and \eqref{flowgredu} are extremely similar to those of the two-dimensional sine-Gordon model with the coupling $u_k$ 
associated to the two-derivative term (and with the four-derivative term neglected). The essential qualitative difference is the negative sign  of
 the right hand side of Eq. \eqref{flowuredu} which is instead positive in the two-dimensional case; conversely,  Eq. \eqref{flowgredu} has the same 
 structure in both cases.
 
\begin{figure}[ht!]
\centering
\includegraphics[scale=.35]{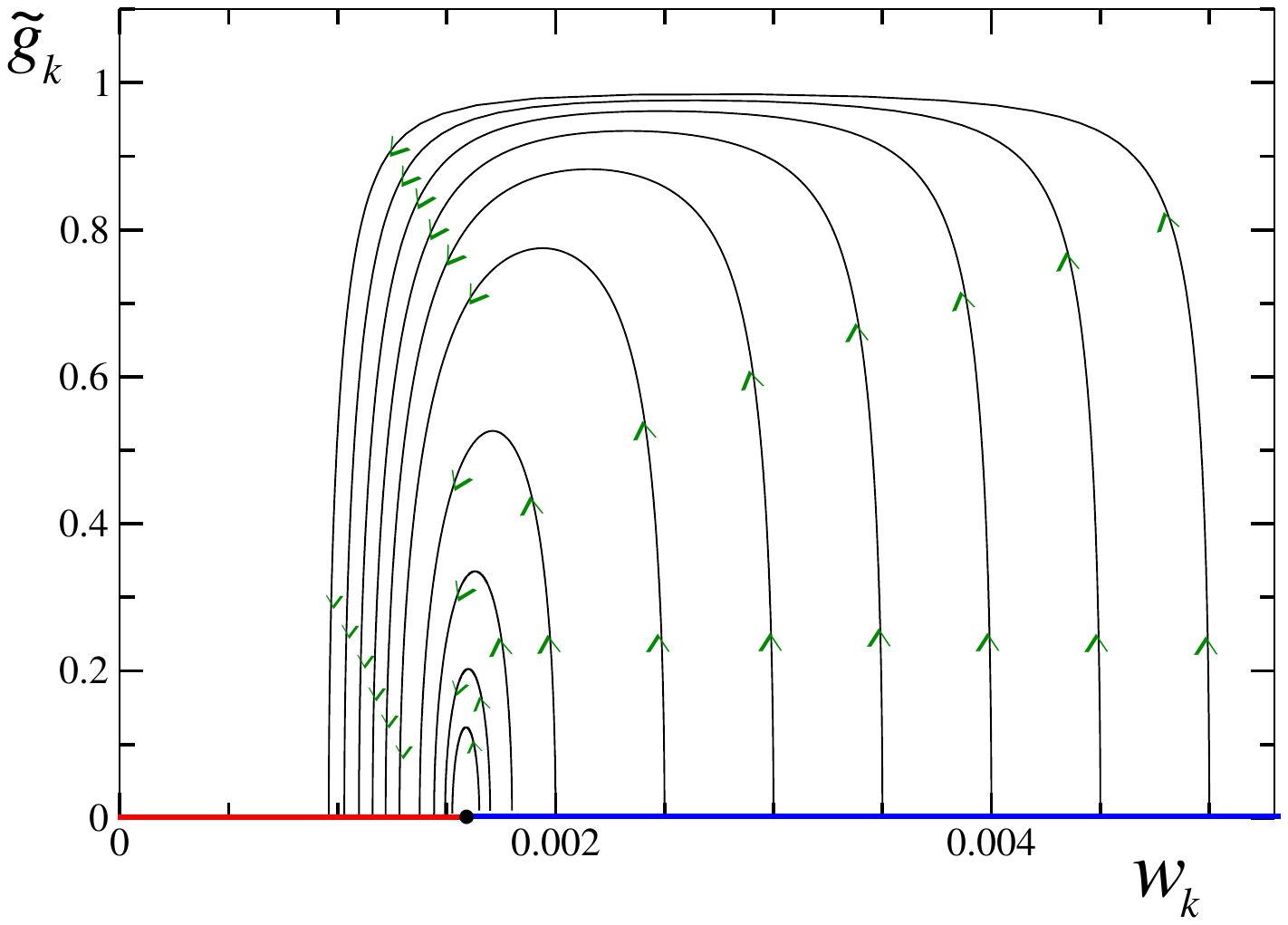}
\caption{Structure of the flow associated to Eqs. \eqref{flowuredu} and \eqref{flowgredu}. 
$\widetilde g_k =0$ is a fixed point  line,  with (blue online) repulsive points above $u^*_k={1}/({64\pi^2})$ and  (red) attractive points below
$u^*_k$  .}
\label{Fig1}
\end{figure}

 As a consequence, the two and four-dimensional Sine Gordon model share the same properties on the $\widetilde g_k =0$ line,
 as it can be checked by first expanding Eq. \eqref{flowgredu}  for small $\widetilde g_k \simeq 0$,   and then taking the limit $\widetilde g_k \to 0$.
In this limit one finds  $\partial_t \widetilde g_k \to  0^+$ and $\partial_t \widetilde g_k \to  0^-$ respectively for $u_k>1/(64\pi^2)$ and $u_k<1/(64\pi^2)$,
and in both cases $\partial_t u_k =  0$. Then one concludes that $\widetilde g_k =0$  is a line of fixed points for the coupled equations
\eqref{flowuredu} and \eqref{flowgredu}, that are repulsive or attractive (when the scale flows toward the IR $k\to 0$), respectively above and below the point
\begin{equation}
\label{critpoint}
u^*_k=\frac{1}{64\pi^2} \; .
\end{equation} 
 
The differences between the two and four-dimensional Sine Gordon model arise when 
$\widetilde g_k \neq 0$, because of the opposite sign in Eq. \eqref{flowuredu}. 
In fact, for the two-dimensional case the positive sign produces the well known BKT phase structure with one phase characterised by the IR  attractive fixed points 
below $u^*_k$ separated by the other phase which is instead dominated by the repulsive fixed points above $u^*_k$. In this case the negative sign forces 
$u_k$ to decrease whenever $\widetilde g_k \neq 0$ and in Fig. \ref{Fig1} the structure of the flow 
generated by Eqs. \eqref{flowuredu} and \eqref{flowgredu} 
is shown. As expected, all lines terminate at one IR fixed point with  $u_k< u^*_k$ and $\widetilde g_k =0$, on the red segment. 
Points with $u_k> u^*_k$ and $\widetilde g_k =0$, are unstable fixed points. 
However, for  $\widetilde g_k \neq 0$, the flow in Fig. \ref{Fig1}  is uniform and no line separating different regions of the diagram, 
i.e. no  phase structure is present:  the system shows only one phase on the plane  $(u_k-\widetilde g_k)$.
 
 As a double check of this neat difference with respect to the two-dimensional case, in Appendix \ref{appendixb} we derive the flow equations by following
 a perturbative approach developed in \cite{Jose1977} for the two dimensional Coulomb gas. In fact, it turns out that the negative sign in 
 Eq. \eqref{flowuredu}  is due to the specific number of dimensions, $d=4$, analysed in this case.
Nevertheless, the above picture is not complete, as we have totally neglected the effect of the relevant coupling $\widetilde z$ so far.
In the next two Sections we show the important role of $\widetilde z$ in establishing the phase structure of model \eqref{model}.

\section{Truncated flow equations}
\label{expanded}

Instead of tackling the full flow in Eqs. \eqref{rgsys}, we start by considering a simplified problem where the differential equations are studied 
around the axis $\widetilde g_k =\widetilde z_k =0$ which, as we will see, is still a line of fixed points, and therefore we expand the right-hand side of 
Eqs. \eqref{rgsys} in powers of $\widetilde g_k$ and $\widetilde z_k $ and retain terms  at most  quadratic in the product of these two parameters.
The output of the expansion is:

\begin{subequations}
\label{linear}
\begin{align}\
\partial_tu_k&=-\frac{\widetilde{g}_k^2}{160\pi^2}\label{linearizedflowa}\\
\partial_t\widetilde{z}_k&=2\widetilde{z}_k+\frac{3}{512\pi\sqrt{u_k}}\widetilde{g}_k^2\label{linearizedflowb}\\
\partial_t\widetilde{g}_k&=(4-\frac{1}{16\pi^2u_k})\widetilde{g}_k+\frac{\widetilde{g}_k\widetilde{z}_k}{64\pi u_k^{3/2}}
\label{linearizedflowc}
\end{align}
\end{subequations}

First we notice that  Eqs. \eqref{linear} are invariant under the transformation 
$\widetilde g_k \to - \widetilde g_k$ and therefore the findings with  $\widetilde g_k>0$, discussed below,
hold also if the sign of the coupling $\widetilde g_k$ is flipped. 
Then, a  direct analysis of the zeroes of the truncated  differential equations,
shows that the only fixed points solutions of Eqs. \eqref{linear} are, as already anticipated,
all the points of  the axis  $\widetilde g_k =\widetilde z_k =0$  (we focus only on positive $u_k>0$, in order to have a stable model).
On the other hand, a direct inspection of Eq. \eqref{linearizedflowc} shows that the point $u_k^*$ in Eq. \eqref{critpoint}
still represents the critical point that marks the different nature of the fixed points  for $u_k <u_k^*$ or $u_k>u_k^*$. 
In fact 
$\widetilde z_k$ is clearly a relevant coupling as evident from Eq. \eqref{linearizedflowb}, then, for $u_k>u_k^*$, 
all points on this axis  are repulsive fixed points all trajectories flow away from them, while for $u_k<u_k^*$,
unlike the diagram  in Fig. \ref{Fig1} there is one new relevant direction associated to $\widetilde z_k$  and the points are no longer IR stable.

Therefore, one concludes  that all points characterized by $\widetilde z_k=0$ and $\widetilde g_k \neq 0$ are not fixed points:
even the single Eq. \eqref{linearizedflowb} has a non-vanishing right-hand side at these points, i.e $\partial_t\widetilde{z}_k \neq 0$.
This means that we cannot expect to observe the flow depicted in Fig. \ref{Fig1} when  the complete set  of equations
\eqref{linear} is solved with initial conditions taken on the plane  $\widetilde z_k=0$ (with $\widetilde g_k \neq 0$),
i.e. we cannot expect the plane  $\widetilde z_k=0$ to be the manifold that separates two different phases of our model, because  it is crossed 
by the  trajectories generated by equations \eqref{linear}. 
Nevertheless, we can still check the behavior of the flow around the two manifolds defined by the two constraints 
$\partial_t\widetilde{z}_k=0$  and $\partial_t\widetilde{g}_k=0$.

\begin{figure}
\centering
\includegraphics[width=0.72\columnwidth]{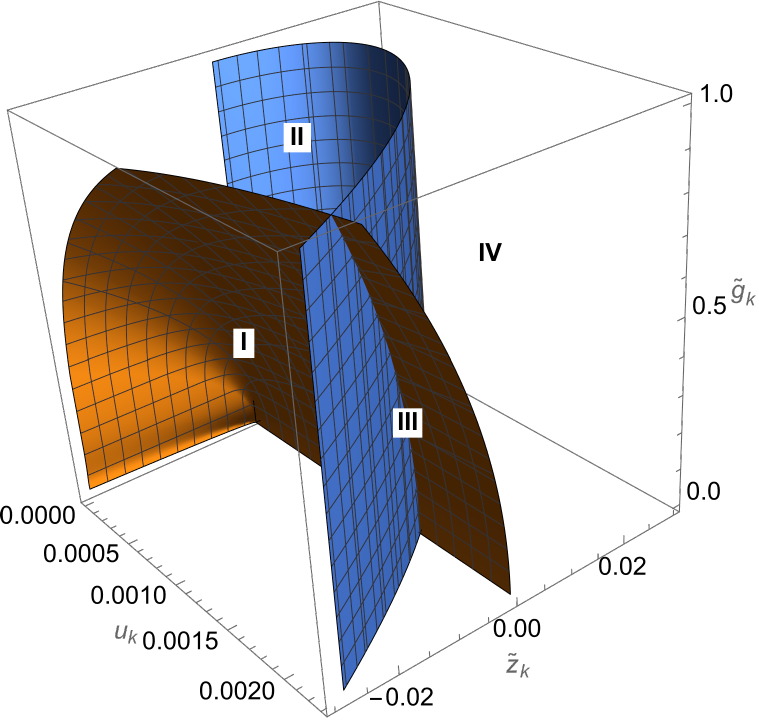}
\caption{Three-dimensional  representation of the manifolds defined by  $\partial_t\widetilde{z}_k=0$ which embeds the line of fixed points
$\widetilde g_k =\widetilde z_k =0$  (orange online) and $\partial_t\widetilde{g}_k=0$ (blue online). See text for the description of regions 
I, II, III, IV.}
\label{fig:superfici}
\end{figure}

These manifolds are plotted in Fig.~\ref{fig:superfici}. Namely, the surface corresponding to $\partial_t\widetilde{z}_k=0$ is the one that 
embeds the straight line of fixed points $\widetilde g_k =\widetilde z_k =0$  (orange surface online),
while  the other manifold  corresponds to $\partial_t\widetilde{g}_k=0$ (blue online).
More precisely, in addition to the points on the  latter manifold, 
the constraint $\partial_t\widetilde{g}_k=0$  also holds for all points on the plane  $\widetilde{g}_k=0$.

The two manifolds  in Fig.~\ref{fig:superfici} separate four different regions: 
region I,  where $\partial_t\widetilde{z}_k<0$, $\partial_t\widetilde{g}_k<0$;
 region II, where $\partial_t\widetilde{z}_k>0$, $\partial_t\widetilde{g}_k<0$; 
 region III, where $\partial_t\widetilde{z}_k<0$, $\partial_t\widetilde{g}_k>0$;
 region IV, where $\partial_t\widetilde{z}_k>0$, $\partial_t\widetilde{g}_k>0$.
 Then, one can predict the trajectory of the flow, according to the location  of the starting point.

\begin{figure}[!h]   
    \centering
    \subfloat[][\emph{$u_k-\widetilde z_k$ plane}]{\includegraphics[width=0.5\textwidth]{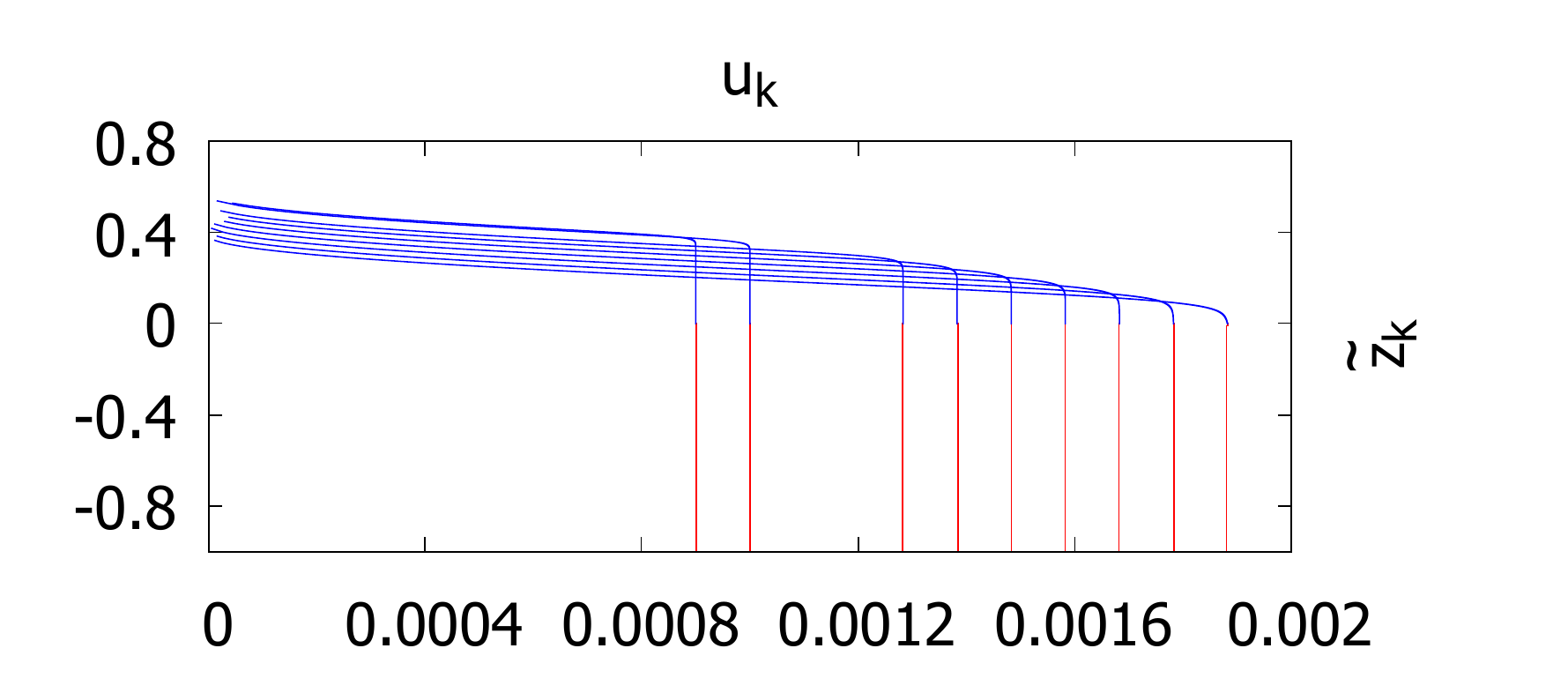}}
    \subfloat[][\emph{$u_k-\widetilde g_k$ plane}]{\includegraphics[width=0.5\textwidth]{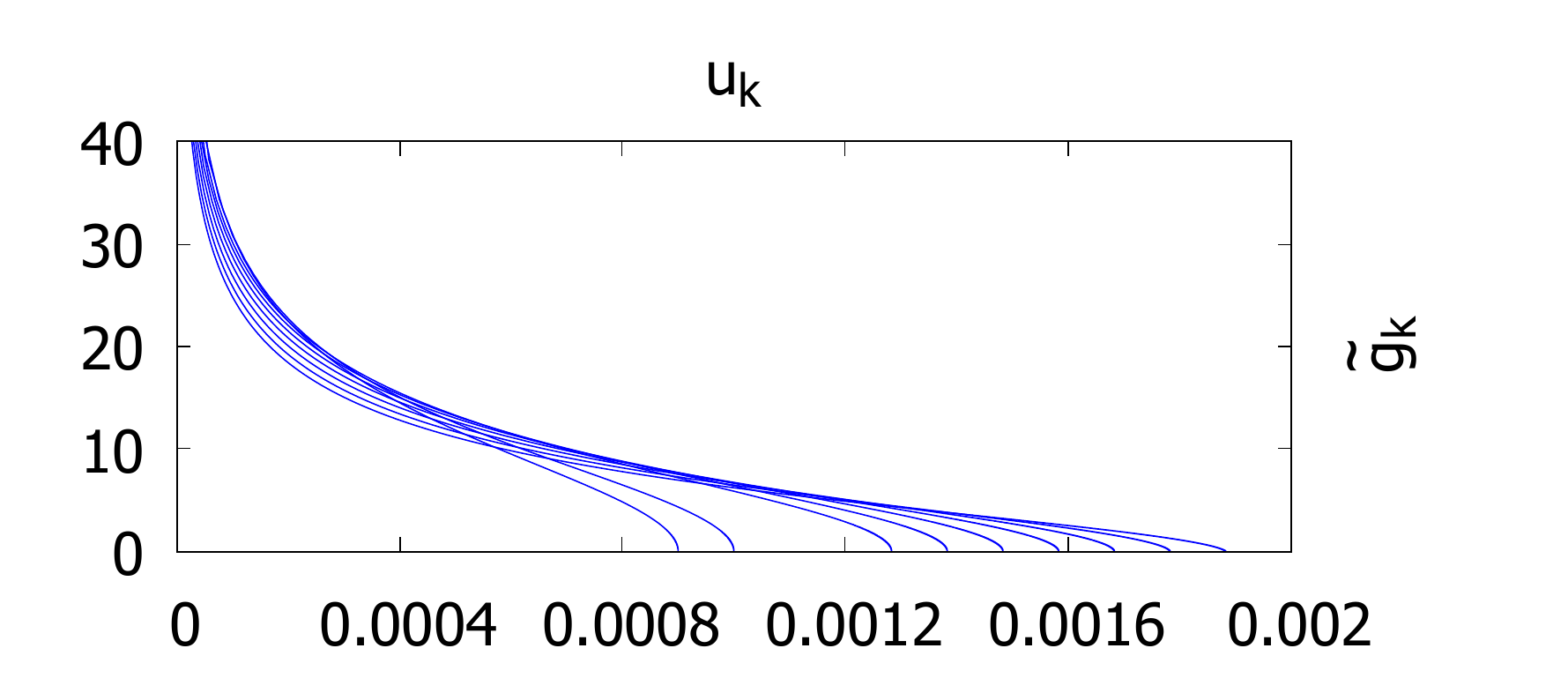}}\\
    \subfloat[][\emph{$\widetilde z_k - \widetilde g_k$ plane}]{\includegraphics[width=0.5\textwidth]{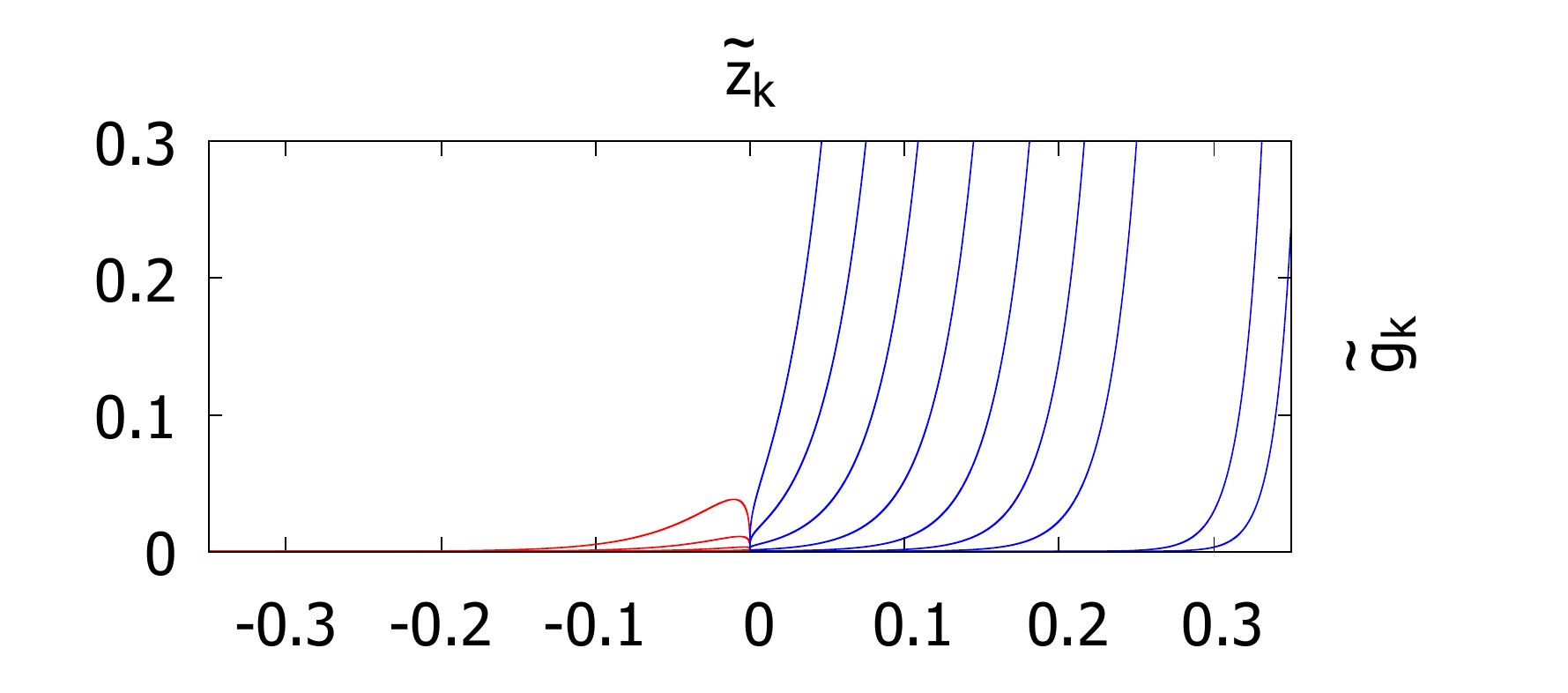}}
       \caption{Projections of the flow derived from Eqs. \eqref{linear}, on the three coordinate planes associated with  $u_k$, $\widetilde z_k$, $\widetilde g_k$.
       The initial points have  different values of $u_k$, but $\widetilde{g}_k=10^{-3}$ for all lines,  while $\widetilde{z}_k=-10^{-7}$ for lines 
    that flow toward large negative $\widetilde z_k$ (red lines for online figures)  and $\widetilde{z}_k=0$ for those that flow toward positive
     $\widetilde z_k$  (blue lines for online figures). Note that in plot (b) only the latter blue curves with positive $\widetilde z_k$ are visible
      due to the scale used on the $\widetilde g_k$ axis.}
    \label{fig:g=10^-3}
    \centering
    \subfloat[][\emph{$u_k-\widetilde z_k$ plane}]{\includegraphics[width=0.5\textwidth]{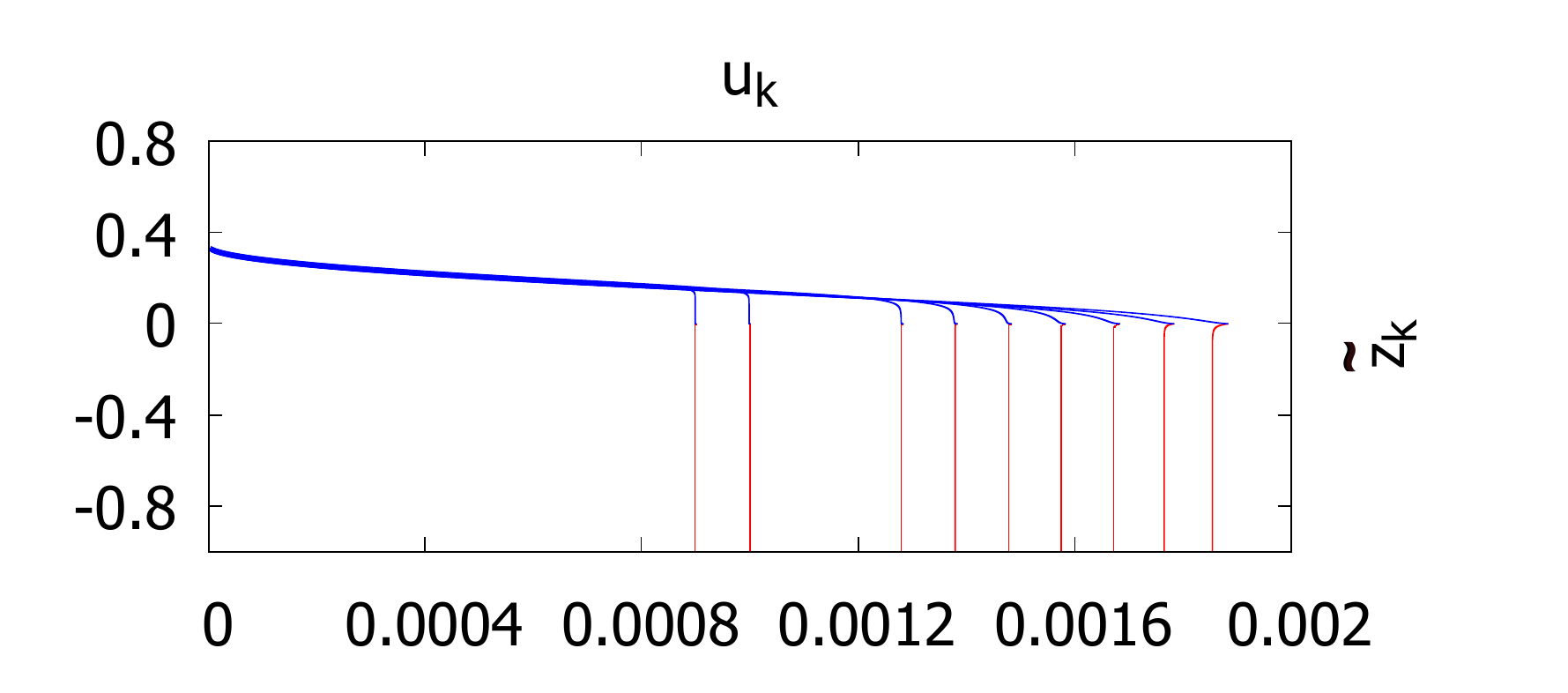}}
    \subfloat[][\emph{$u_k-\widetilde g_k$ plane}]{\includegraphics[width=0.5\textwidth]{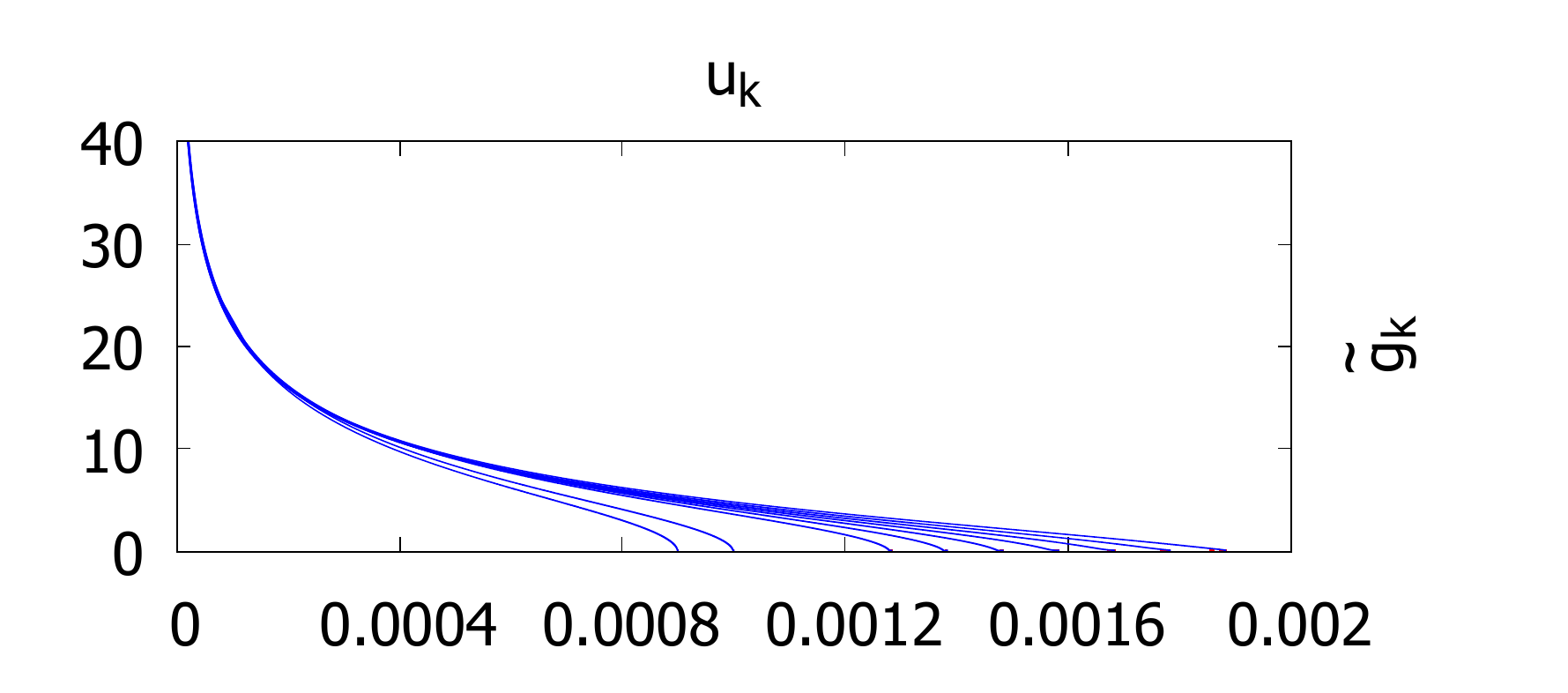}}\\
    \subfloat[][\emph{$\widetilde z_k - \widetilde g_k$ plane}]{\includegraphics[width=0.5\textwidth]{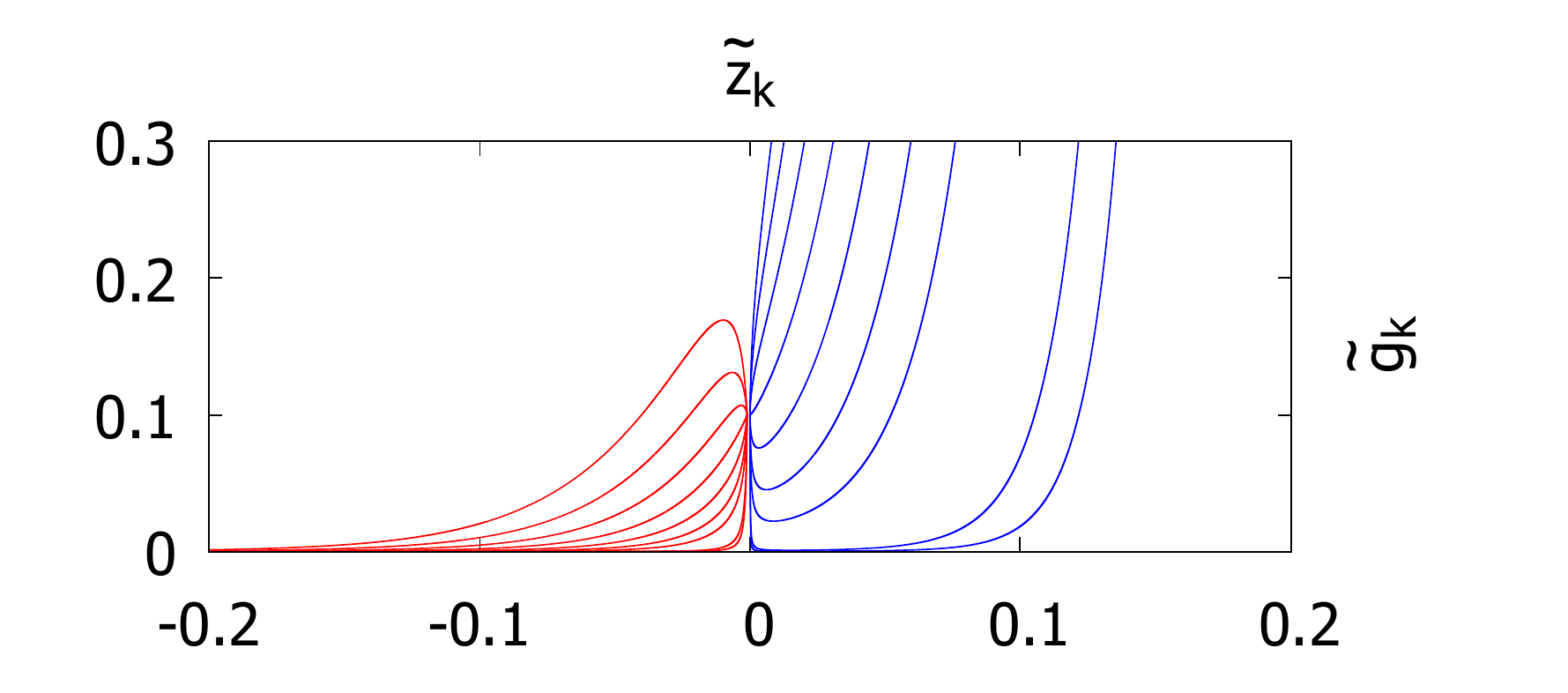}}
    \caption{Same as in  Fig. \ref{fig:g=10^-3}, but  $\widetilde{g}_k=10^{-1}$ in all cases,
                  while $\widetilde{z}_k=-10^{-3}$ for lines that flow toward large negative $\widetilde z_k$ (red curves for online figures)
                  and $\widetilde{z}_k=-10^{-4}$ for those that flow toward positive  $\widetilde z_k$  (blue curves for online figures).
                  As in Fig. \ref{fig:g=10^-3}, in plot (b) only the latter blue curves with positive $\widetilde z_k$ are visible due to the
                   scale used on the $\widetilde g_k$ axis. } 
\label{fig:g=10^-1}
\end{figure}

\begin{itemize}
\item[1)] The initial point is taken in region I. In this case  $\widetilde{z}_k$ gets more and more  negative  and $\widetilde{g}_k$ decreases to $0$ safely.
Note that the possibility that the trajectory crosses the plane $\widetilde{g}_k=0$  is excluded because on this plane $\partial_t\widetilde{g}_k=0$ 
 so the trajectory is forced to have $\widetilde{g}_k>0$.  
 \item[2)] The initial point is taken in region IV.  $\widetilde{z}_k$ and  $\widetilde{g}_k$ grow positive but  $u_k$ decreases until it reaches  $0$, which is a spinodal point 
as Eqs. \eqref{linearizedflowb} and \eqref{linearizedflowc} are singular at $u_k=0$.
\item[3)] The initial point is located within region II. In this case one can distinguish  two cases. In fact, either  $\widetilde{z}_k$ grows 
 positive while $\widetilde{g}_k$ decreases and the trajectory   crosses  the orange surface and the flow afterwards proceeds according to  point 1), 
 or, if the the initial point is too far from the orange manifold the trajectory  crosses  the blue manifold and the conditions of point  2) are recovered.
 As in point 1),  the trajectory cannot cross  the plane $\widetilde{g}_k=0$   and therefore  $\widetilde{g}_k$ remains positive 
 until  the blue manifold is eventually crossed.
 \item[4)] The initial point is taken in region III. In this case,  $\widetilde{g}_k$  grows and  $\widetilde{z}_k$  diminishes so that 
the orange surface is crossed and region IV is entered and we are back to point 2), unless the initial point is very close to the blue 
manifold, so that the trajectory crosses the latter and ends up in region I, as in the case of point 1).
\end{itemize}

Finally, if the initial point is taken on the plane $\widetilde{g}_k=0$,  both $u_k$  and $\widetilde{g}_k$ remain unchanged, according to 
Eqs. \eqref{linear}, and the flow concerns $\widetilde{z}_k$ that grows to large positive (negative) values 
if its initial value is positive (negative). Then, in all cases the trajectories  in the IR limit end either in region I or  region IV.

This analysis clearly signals the presence of two phases associated to the different behavior of the trajectories  in the IR limit,
and one can speculate on the features of the manifold that represents the separatrix of the flows in the two phases. In fact,
this manifold must contain the straight line of fixed points $\widetilde g_k =\widetilde z_k =0$, and, in proximity of this line, it has to stay very close to 
the plane $\widetilde z_k =0$,  although for larger $\widetilde g_k$ it is bent toward negative $\widetilde z_k$. An initial point taken 
on the separatrix generates a flow totally contained within this manifold, that in the IR limit reaches a fixed point on the line with $\widetilde g_k =\widetilde z_k =0$
 and $u_k<u_k^*$, in a way that resembles the flow in Fig. \ref{Fig1}, where  $\widetilde z_k =0$ is artificially enforced.

To visualize the separation of the  flows in the two phases, we show below the plots of the  trajectories that start  at fixed $\widetilde g_k$ with 
two slightly different values of $\widetilde z_k$, located on different  sides of the separatrix, thus producing flows that end either in region I or IV.
Namely in Fig. \ref{fig:g=10^-3}, for nine different initial values of $u_k$, we choose  $\widetilde g_k =10^{-3}$, and $\widetilde{z}_k=-10^{-7}$ for one 
phase  and $\widetilde{z}_k=0$ for the other; in Fig. \ref{fig:g=10^-1} for the 
same  $u_k$, we take $\widetilde g_k =10^{-1}$, and $\widetilde{z}_k=-10^{-3}$ for one phase  and 
$\widetilde{z}_k=-10^{-4}$  for the other. 

In both figures, we report the projection of the flow on the three coordinate planes associated with $u_k$,
$\widetilde z_k$, $\widetilde g_k$. The different behavior of the flow in the two phases is evident  both in the (a)  and (c)  plots of Figs. \ref{fig:g=10^-3},
 \ref{fig:g=10^-1}. In fact, in one phase (red curves for online figures),  $\widetilde{z}_k$ gets more and more negative, $u_k$ stays constant and 
$\widetilde{g}_k$, after a small bump, rapidly drops to zero. 
In the other phase (blue curves for online figures), $\widetilde{z}_k$ becomes positive while 
$u_k$ approaches zero and, at the same time, $\widetilde{g}_k$ rapidly grows, so that the flow ends by hitting a spinodal point at $u_k=0$. In the (b) plots
the red curves with negative $\widetilde{z}_k$ of the former phase, are invisible due to the large scale adopted to plot the large growth of  
$\widetilde{g}_k$. Finally we notice that, going from Figs. \ref{fig:g=10^-3} to Fig. \ref{fig:g=10^-1}, no qualitative change in the flow is observed,
but only more pronounced changes of the variables $u_k$ and $\widetilde{g}_k$.

\section{Complete flow equations}
\label{complete}

Once we have a definite picture of the truncated flow in Eqs. \eqref{linear} which is certainly reliable for small couplings $\widetilde z_k$, $\widetilde g_k$,
we switch to the study of  the complete RG  beta-functions \eqref{rgsys}, whose  explicit form is displayed in   Eqs. \eqref{flowu}, \eqref{flowz},  \eqref{flowg}.
Even if not evident from these three equations, properties observed in the truncated case in \eqref{linear} for  small $\widetilde g_k$, $\widetilde z_k$
must obviously hold even in the general case in the same regime.

\begin{figure}[!h]
    \centering
    \subfloat[][\emph{$u_k-\widetilde z_k$ plane}]{\includegraphics[width=0.5\textwidth]{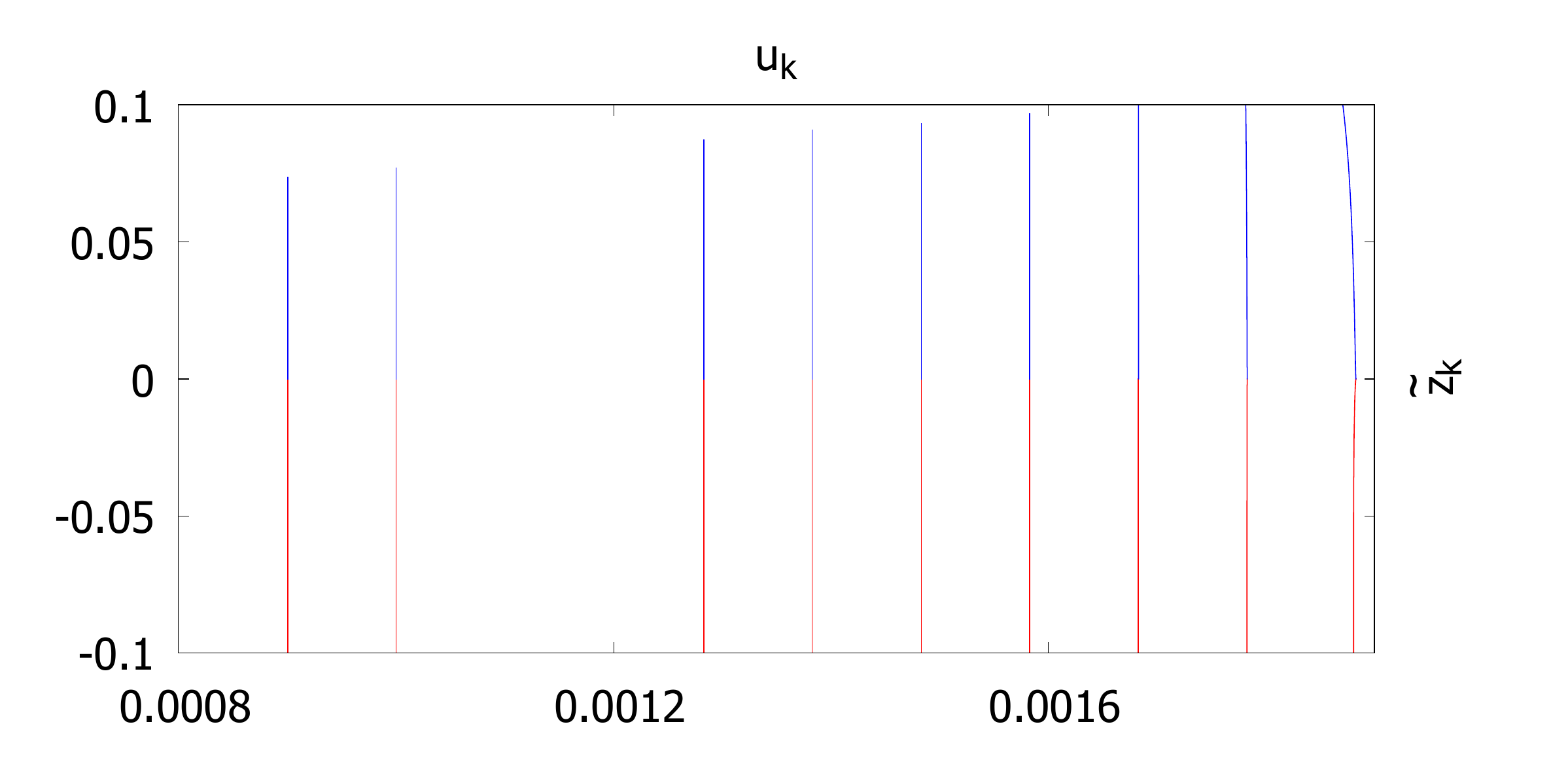}}
    \subfloat[][\emph{$u_k-\widetilde g_k$ plane}]{\includegraphics[width=0.5\textwidth]{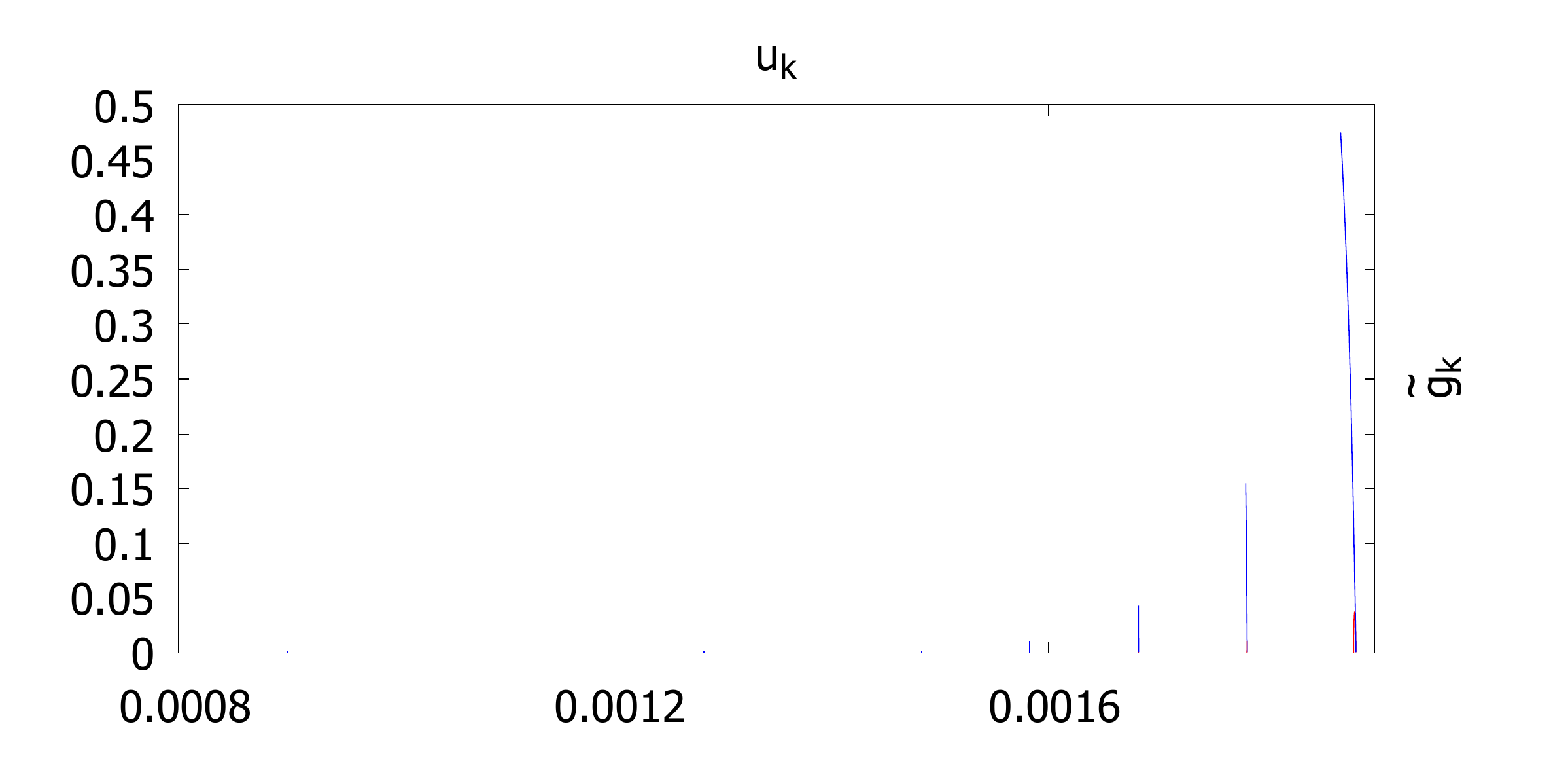}}\\
    \subfloat[][\emph{$\widetilde z_k - \widetilde g_k$ plane}]{\includegraphics[width=0.5\textwidth]{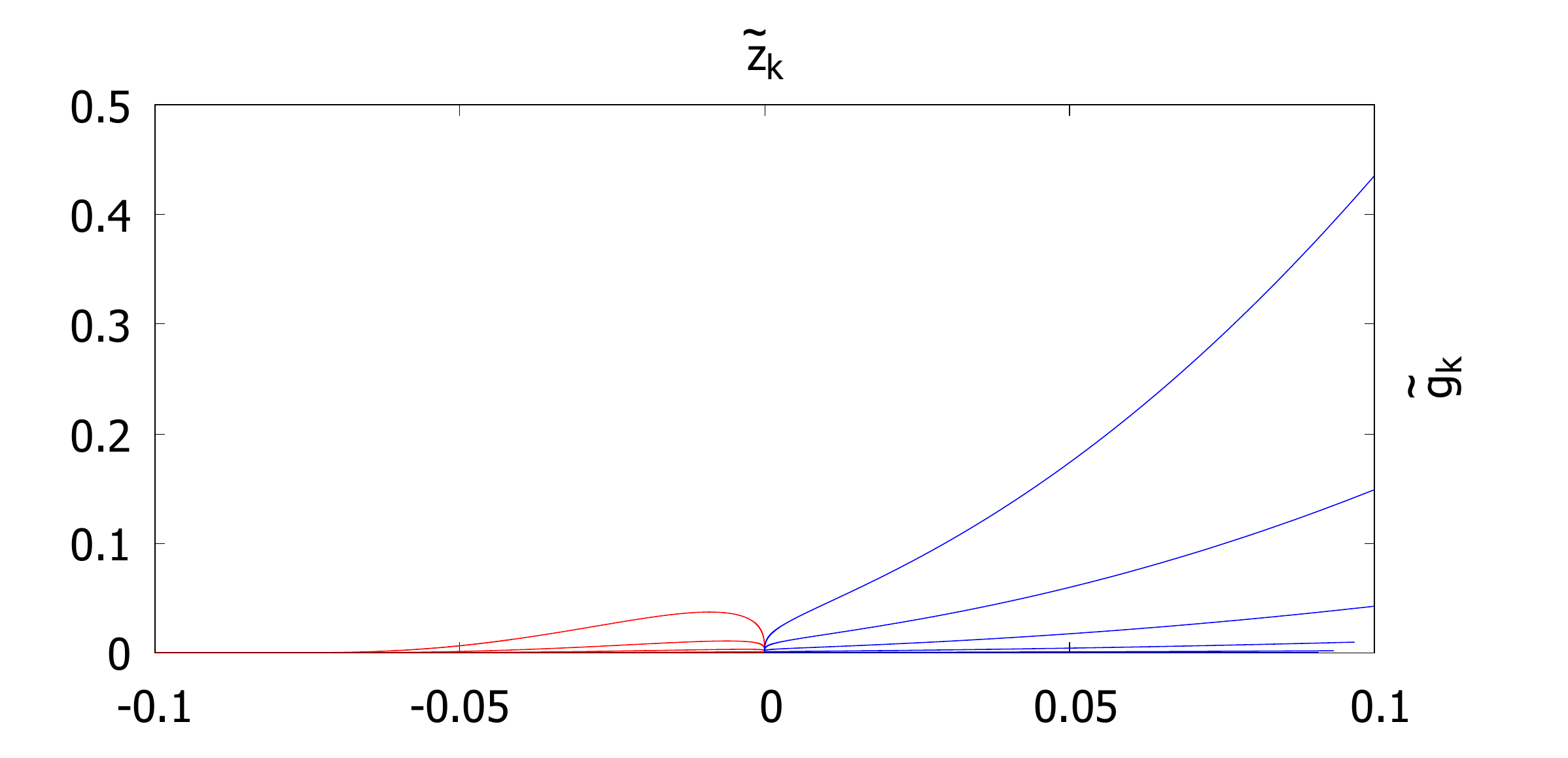}}
    \caption{Projections of the flow derived from Eqs. \eqref{flowu}, \eqref{flowz},  \eqref{flowg},
    on the three coordinate planes  associated with $u_k$, $\widetilde z_k$, $\widetilde g_k$.     
      The initial points have  $\widetilde{g}_k=10^{-3}$ for all lines,  while $\widetilde{z}_k=-10^{-7}$ for lines 
    with $\widetilde z_k<0 $ (red lines for online figures)  and $\widetilde{z}_k=10^{-7}$ for those with
     $\widetilde z_k>0$  (blue lines for online figures). As in Fig. \ref{fig:g=10^-3}, 
      only the latter blue curves with $\widetilde z_k>0$ are visible in plot (b).}
    \label{fig:Full_10^-3}
    \centering
    \subfloat[][\emph{$u_k-\widetilde z_k$ plane}]{\includegraphics[width=0.5\textwidth]{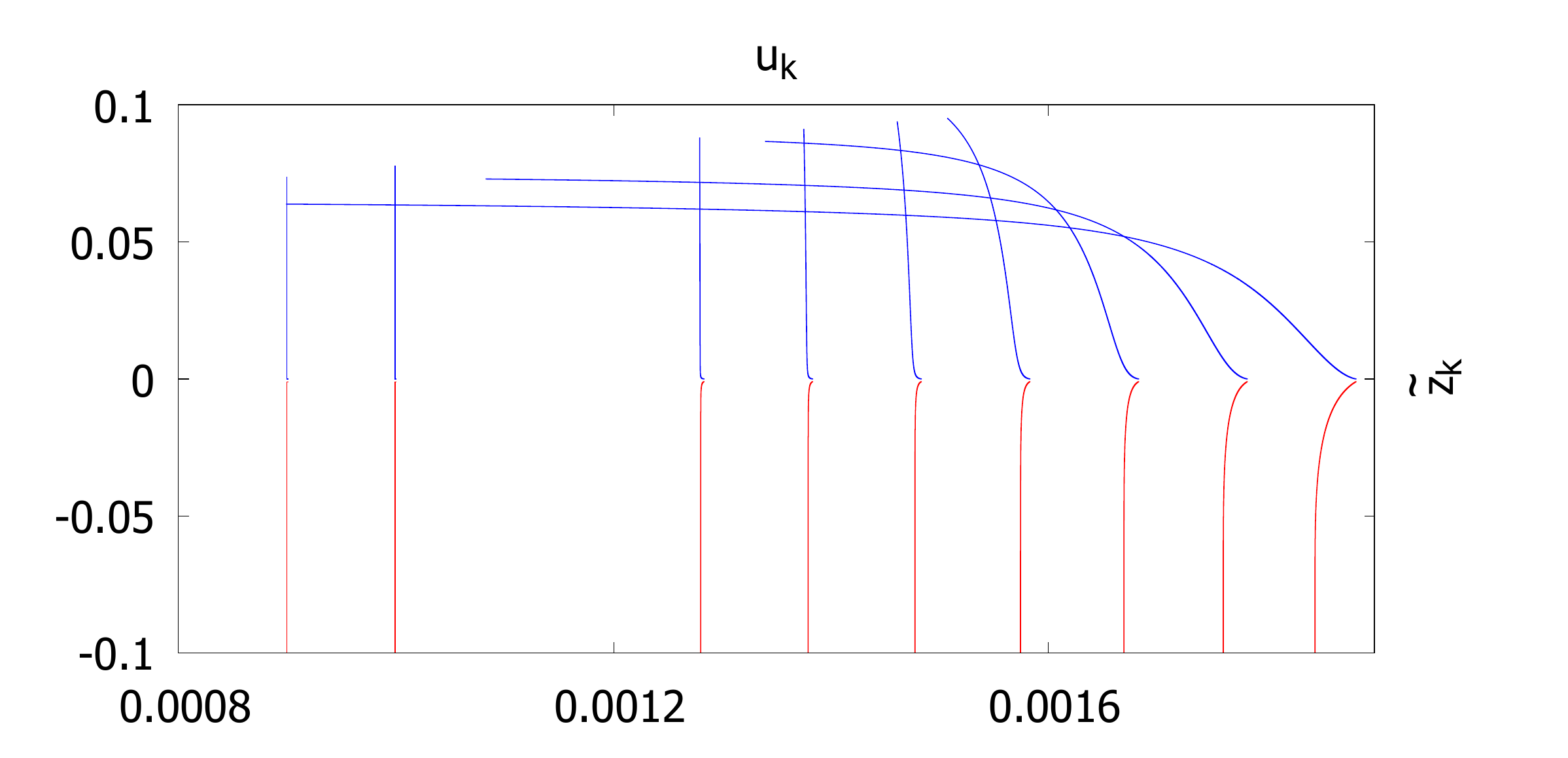}}
    \subfloat[][\emph{$u_k-\widetilde g_k$ plane}]{\includegraphics[width=0.5\textwidth]{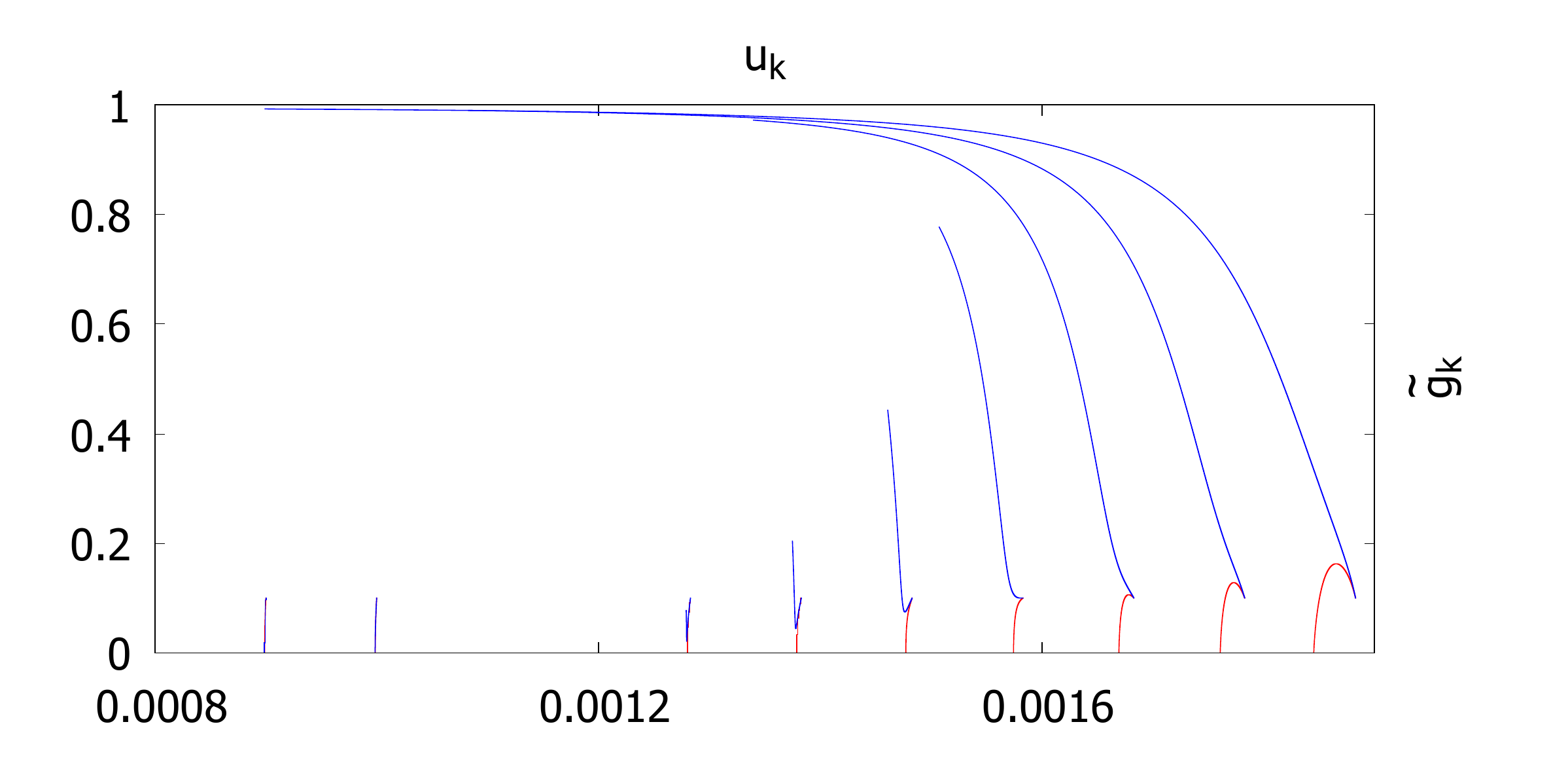}}\\
    \subfloat[][\emph{$\widetilde z_k - \widetilde g_k$ plane}]{\includegraphics[width=0.5\textwidth]{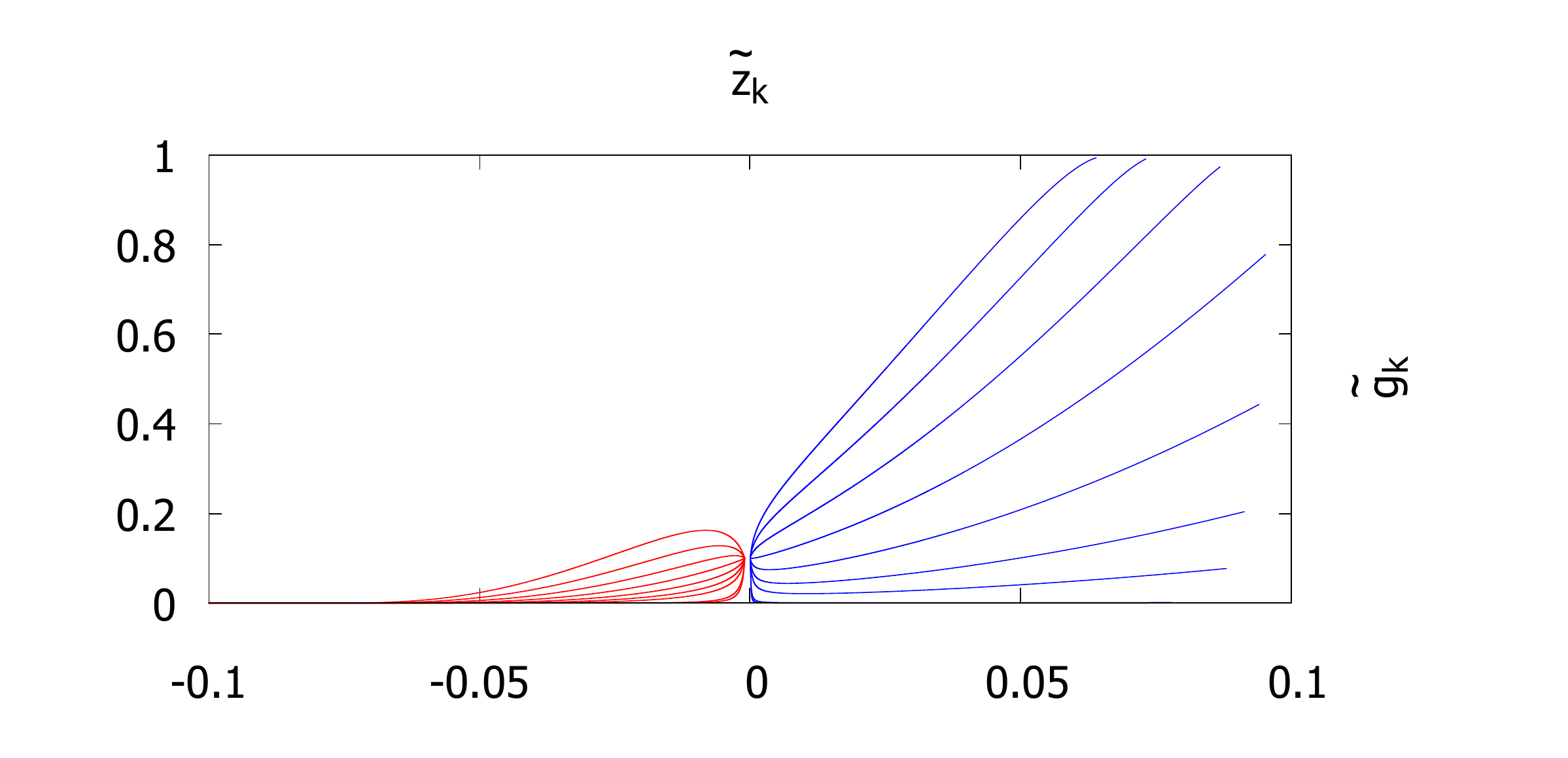}}
    \caption{Same as in Fig. \ref{fig:Full_10^-3}, but with initial values $\widetilde{g}_k=10^{-1}$ for all lines 
     while $\widetilde{z}_k=-10^{-3}$ for curves with $\widetilde z_k<0 $ (red online)    
     and $\widetilde{z}_k=10^{-7}$ for curves with $\widetilde z_k>0 $ (blue online). }
    \label{fig:Full_10^-1}
\end{figure}

Therefore, the line of fixed points at  $\widetilde g_k =\widetilde z_k =0$ is still present in Eqs. \eqref{flowu}, \eqref{flowz},  \eqref{flowg}
and the properties of this line of points, already  discussed for the truncated equations, remain unchanged also when  the complete flow equations are considered.
In addition, close to this axis,  in Figs. \ref{fig:Full_10^-3} and \ref{fig:Full_10^-1}  we can deduce again
the presence of a separatrix manifold between the two phases characterised by positive  (blue curves for online figures) or negative
(red  curves for online figures) $\widetilde z_k$ for growing $t=\log (\Lambda/k)$.

\begin{figure}[h!]
    \centering
\subfloat[][\emph{}]{\includegraphics[width=0.5\textwidth]{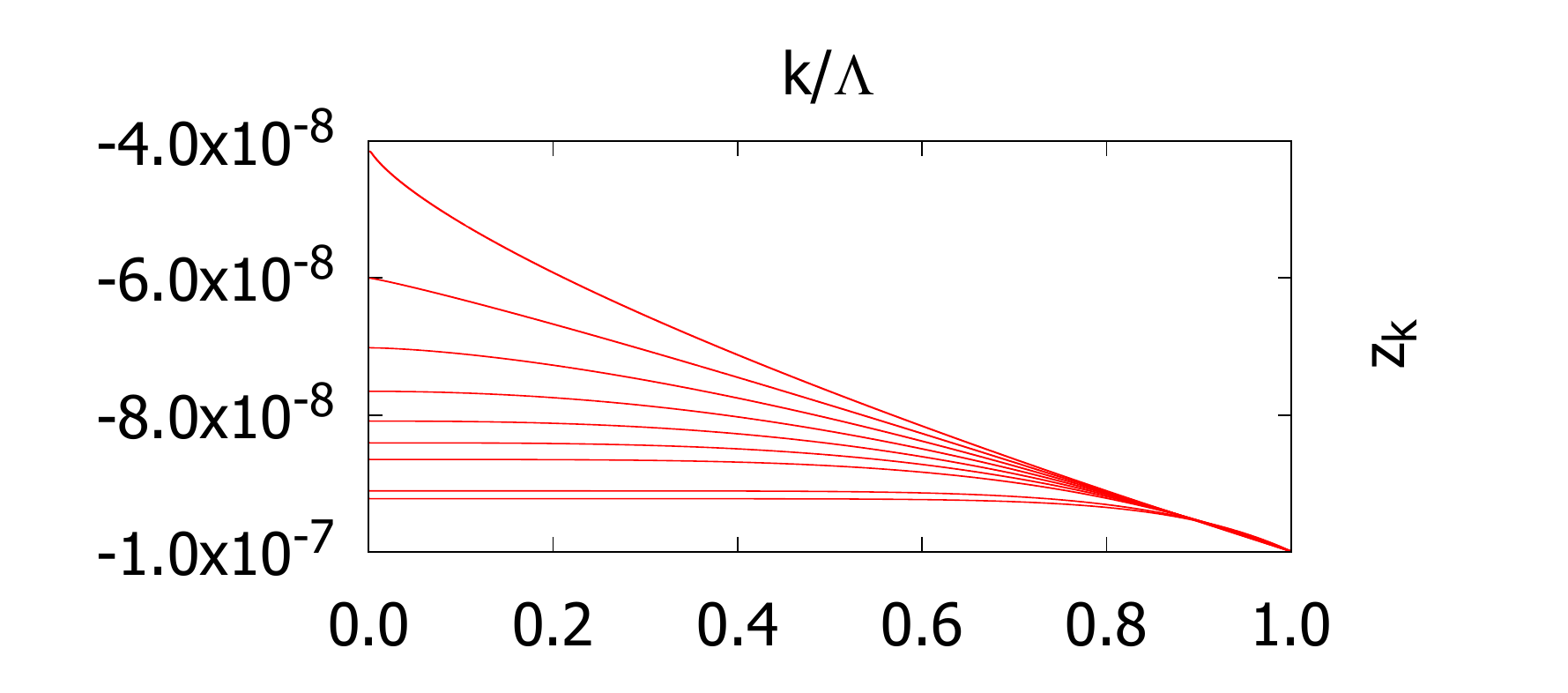}}
\subfloat[][\emph{}]{\includegraphics[width=0.5\textwidth]{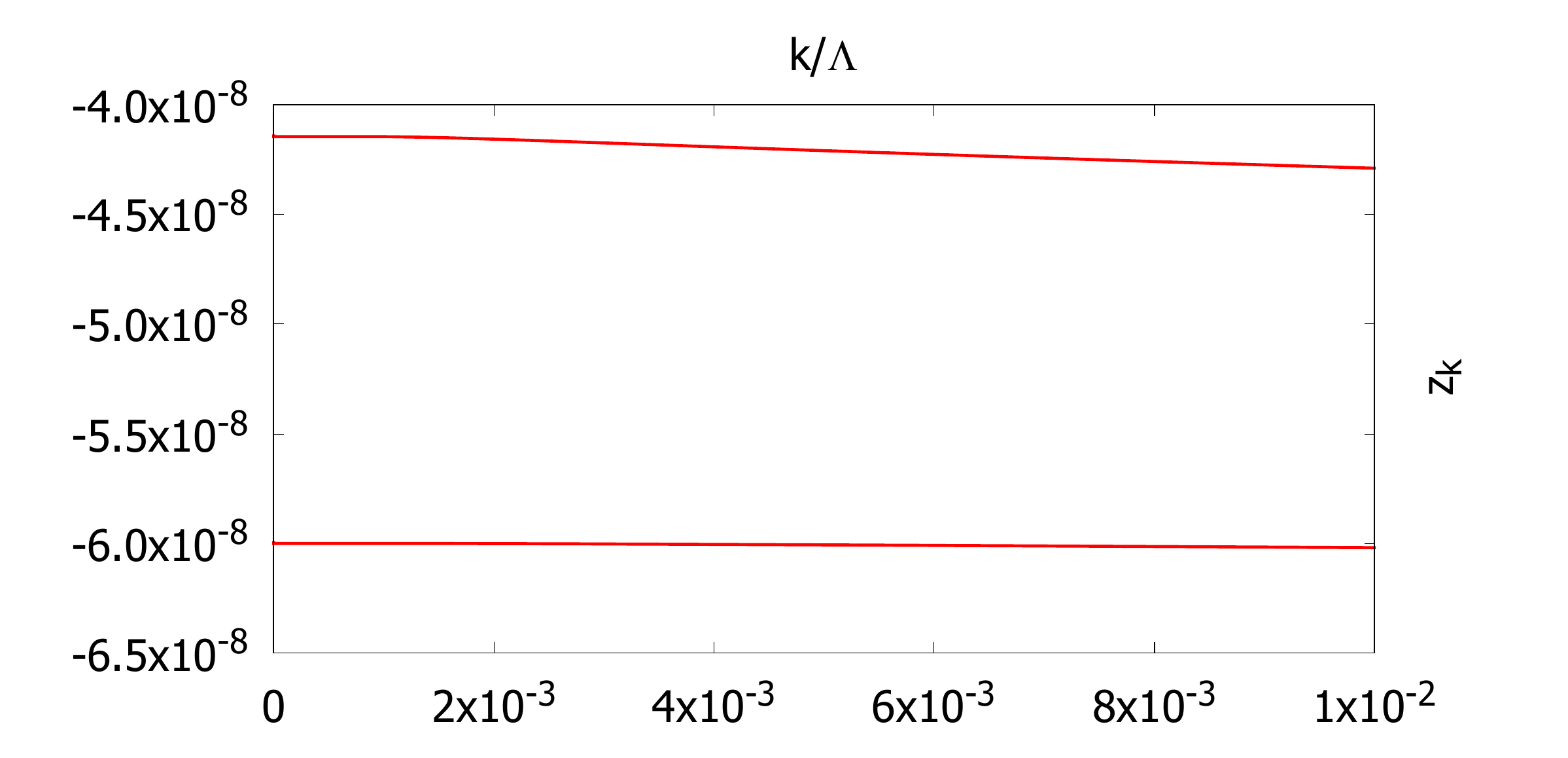}}\\
\subfloat[][\emph{}]{\includegraphics[width=0.5\textwidth]{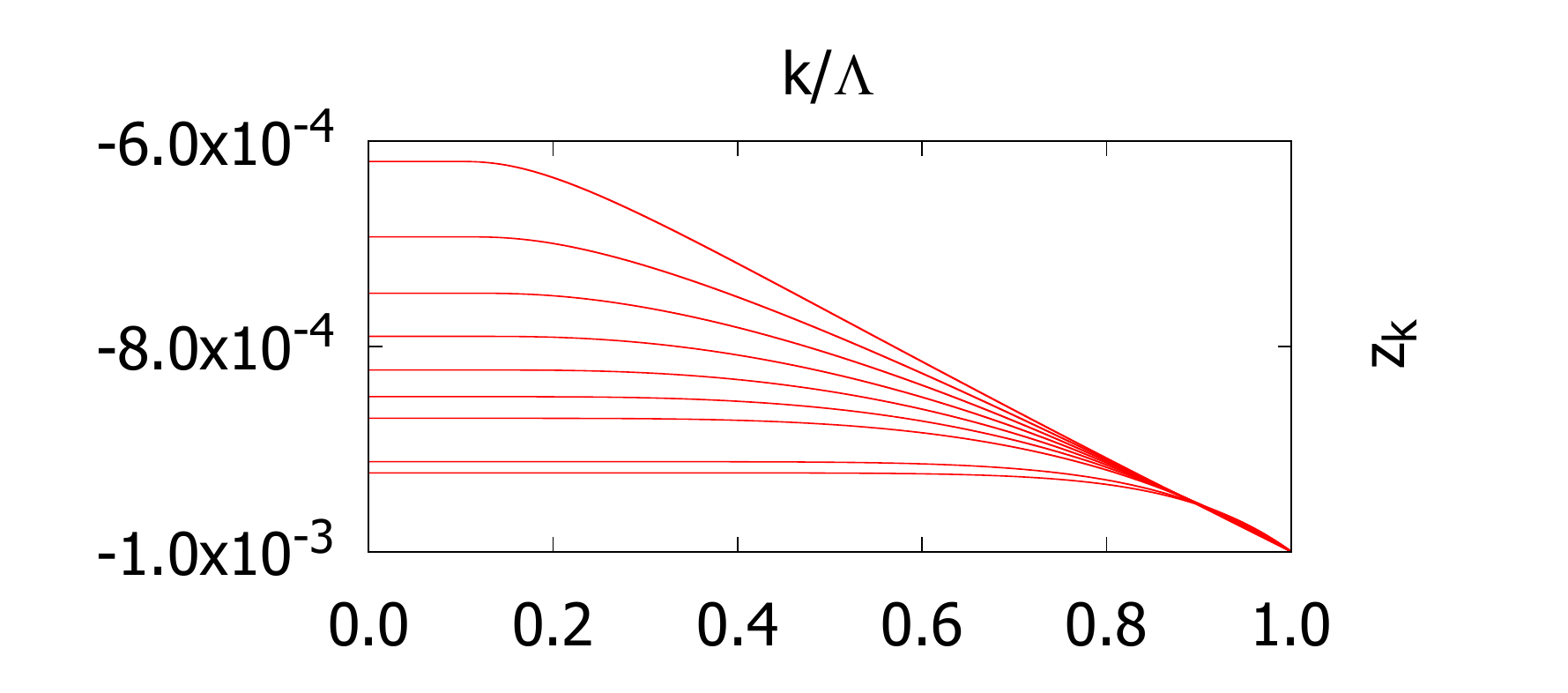}}
\caption{Flow of $z_k=k^2\widetilde{z}_k$.  Lines are plotted accordingly to the coding of the 
     previous figures,  respectively referring to the $\widetilde{z}_k<0$ (red online)  phase in 
      Fig.\ref{fig:Full_10^-3}  (plot (a)) and  Fig.\ref{fig:Full_10^-1}  (plot (c)). Plot (b) shows the enlargement of a detail of plot (a).}
       \label{z_k}
\end{figure}

As for the truncated flow, by comparing 
Figs. \ref{fig:Full_10^-3} and \ref{fig:Full_10^-1}, we see again a swifter evolution of the three parameters in the latter case, i.e.
for larger initial value of $\widetilde g_k$. So, in this regime, no qualitative difference appears with respect to Eqs. \eqref{linear}.

However, a clear difference is evident in the (blue) phase with  positive  $\widetilde z_k$, when confronting Figs.\ref{fig:Full_10^-3},\ref{fig:Full_10^-1} 
with Figs.\ref{fig:g=10^-3},\ref{fig:g=10^-1}. In fact in the latter case, $\widetilde g_k$ grows to very  large values  while $u_k$ is constantly decreasing
and the curves are interrupted only  when they hit a spinodal point with $u_k=0$ and, consequently, $\partial_t \widetilde g_k=\infty$; 
in the complete flow instead,
a weak  dependence of $u_k$ on the flow, especially for smaller values of $\widetilde g_k$ (see Fig. \ref{fig:Full_10^-3}), and, 
more importantly, the curves 
get stopped at points that have rather small values of $\widetilde g_k$ and $\widetilde z_k$, well before  $u_k$  could reach zero.

The interruption  of the curves in the phase with positive  $\widetilde z_k$ is  possibly due
 to a breakdown of the numerical procedure adopted to integrate the equations 
because the exponential factors appearing  in Eqs. \eqref{flowu}, \eqref{flowz},  \eqref{flowg} become too large to be handled. As it is visible especially in
plot (a) and (c) of  Figs.\ref{fig:Full_10^-3},\ref{fig:Full_10^-1}, this occurs approximately for $\widetilde z_k \simeq 0.07 \div 0.1$, depending on the value of
$\widetilde g_k$. Actually, we did not attempt to solve or avoid this  numerical problem, as in any case we observe that the flow in this regime yields increasing 
$\widetilde g_k$ and decreasing $u_k$ and  it is evident, at least from  Eq.  \eqref{flowg},  that  values $\widetilde g_k> 1$ produce complex  numbers in the 
right-hand side of the equation, while $u_k=0$ corresponds to a spinodal point where  $\partial_t \widetilde g_k$ diverges. Therefore, 
the  set of flow equations \eqref{flowu}, \eqref{flowz},  \eqref{flowg}, in the phase  with positive  $\widetilde z_k$  is anyway plagued by singularities.

Instead, we now focus on the other phase with  $\widetilde z_k<0$, which is much smoother and the flow trajectories are easily prolonged down to $k\to 0$ 
with $\widetilde g_k$  that gets rapidly suppressed in that limit, while $u_k$  stays practically unchanged and $\widetilde z_k \to - \infty$.
It is interesting to check the behavior of $g_k$ and $z_k$, related to $\widetilde g_k$ and $\widetilde z_k$ by Eq. \eqref{tilde}. It is easy to verify
that $g_k \to 0 $, because of Eq. \eqref{tilde}, while $z_k$ requires an explicit computation, and   we report its plot as a function of $k$
for the three flows already examined in  Figs.\ref{fig:Full_10^-3}, \ref{fig:Full_10^-1}, 
respectively in plot (a) and (c) of  Fig. \ref{z_k}. Now it is evident that $z_k$ is finite in the limit $k\to 0$ and it has also become constant when $k$ 
approaches zero, as shown by the formation of plateaus in this limit.  Actually, even the two upper curves in plot (a) have such a plateau, but
it becomes visible only at smaller  $k$ as it is shown in the enlargement displayed  in plot (b):  
the constant value is reached around $k\simeq 2 \cdot 10^{-3}$ for the lower curve and $k\simeq 10^{-3}$ for the upper curve.

These results represent a crucial starting  point  to consider  the  IR limit of the  model  \eqref{gammak} 
and to analyse how it is described from the point of view of  the standard two-derivative four-dimensional sine-Gordon model. 
In fact,  in the latter model  the scaling dimensions of the operators are set by using the two-derivative term as the leading 
reference operator and this clearly  produces a different scaling with respect to the one set in Eq. \eqref{tilde} to determine the flow,
with dimensionless field $\varphi$ and dimensionless $u_k=\widetilde u_k$.

Therefore, we go back to Eq. \eqref{model} and require the field $\Phi$ to have the dimension of an energy; consequently,
to maintain dimensionless the argument of the cosine, $\beta$  must have the dimension of an inverse energy and,
 since it must be RG scale independent, we can use it to set a definite mass scale $M$ that specifies the IR regime of the model :  
 \begin{equation}
 \beta_{_M}=\frac{1}{M}
\label{betam}
\end{equation}
Then  accordingly, the  IR average effective action is
\begin{equation}
\Gamma^{IR}_k[\Phi] = 
\int d^4 {x} \; \Big [ \frac{ w_k }{2} \, \Delta \Phi ({x})
\Delta \Phi ({x}) + \frac{Z_k}{2} \, \partial_\mu \Phi ({x})  \partial_\mu \Phi ({x})  + g_k\, (1- \cos \beta_{_M} \Phi )   \Big ]
\label{gammak2}
\end{equation}
and  the dimension of the two couplings of the derivative terms
in energy units are : $[Z_k] =0$,   $[w_k]=-2 $, while the coupling $g_k$  coincides with that in Eq. \eqref{gammak} and therefore $[g_k] =4$, 
and the rescaled field defined in  Eq.  \eqref{redefine} ,  $\varphi= \beta_{_M} \Phi$, is dimensionless.
 
 In order to make contact between the flow in Eq. \eqref{gammak} and in Eq. \eqref{gammak2}, it is sufficient to rewrite  the field  $\Phi$
 in terms of $\varphi$ in the latter, according to Eq.  \eqref{redefine}, and then equate the two expressions. Thus, we find the relations
 \begin{equation}
\frac {w_k}{ \beta_{_M}^2 } = u_k  = \widetilde u_k
\;\;\; ; \;\;\;\;\;\; 
\frac {Z_k}{ \beta_{_M}^2 } = z_k  = \widetilde z_k \, k^2 \; ,
\label{compare}
\end{equation}
which express the parameters  $w_k$ and $Z_k$ in terms of $\widetilde u_k$ and $\widetilde z_k$
in the regime $k<M$.

From Eq. \eqref{compare} and  Fig.\ref{z_k} we conclude  that $Z_k$  is a constant in the limit $k\to 0$. 
Also $w_k$  is constant in the IR limit, so that, this parameter expressed in units of the running scale $k$ goes like
$ \widetilde w_k \equiv w_k \, k^2= u_k \, k^2 / M^2$, i.e. vanishes in the limit $k\to 0$. 
In addition,  as already observed, both $g_k$ and $\widetilde g_k$ vanish in the same limit and one concludes 
that the IR limit of the model \eqref{gammak}, described in terms of the effective model in Eq. \eqref{gammak2}, 
reduces to the two-derivative kinetic term with a finite multiplicative  coefficient $Z_k$, 
which is nothing else than the Gaussian model, but with the essential difference that in this case $Z_k<0$.

\section{Concluding remarks }
\label{conclusion}
We studied the generalized higher-derivative sine-Gordon model in Eq. \eqref{model},  which includes a four-derivative plus a two-derivative 
kinetic term,  in 4D and we obtained an improved picture of its phase diagram, with respect to the model discussed in \cite{defenu}
where the two-derivative term is omitted.  In \cite{defenu},  it was observed that, despite the similarities with the 2D BKT transition
and  despite the possibility of constructing topological configurations analogous to the 2D vortices, the phase diagram obtained in that case 
shows a significant difference with respect to the BKT one.
 In fact, although a line of fixed points is found, including a critical point  at $u_k=u_k^*$ 
that separates a subset of attractive from  another subset of repulsive fixed points, yet the  negative
sign appearing in the right hand side of one of the flow equations  does make all flow trajectories homogeneous  (as shown in Fig \ref{Fig1}),
 without a distinction of two separate phases, as occurs instead  in the 2D  BKT diagram.

When the full model, Eq. \eqref{model} , is taken into account, it becomes clear that the two-derivative operator proportional to $z_k$, a relevant coupling,
is essential in determining the separation of two phases, which are realised in the IR limit  for different  signs of $\widetilde z_k$, and the line  characterised by 
$\widetilde z_k=\widetilde g_k=0$ and $u_k > u_k^*$ is a set of UV attractive fixed points with all trajectories departing from it when the RG scale is lowered.

In the adopted scheme, the phase displaying $\widetilde z_k>0$ shows some pathologies, such as the presence of spinodal points or regions  
where RG equations become non-analytic, that make the IR limit $k\to 0$  impossible to achieve, and  this suggests that  the model in Eq. \eqref{gammak}
could be insufficient to properly describe the $\widetilde z_k>0$ phase. Conversely, the opposite case with $\widetilde z_k<0$ is totally under control and 
the RG flow generated by the action in Eq.  \eqref{gammak} is regular in the IR limit and, with the help of a mapping on an effective model, built  around the 
Gaussian fixed point,  we verified that our system reduces to a Gaussian model in this limit, but with  negative $\widetilde z_k<0$.

The use of another  effective model to analyse the IR limit, is in fact inspired by the work in \cite{buccio} where this approach is used to discuss 
the flow from an UV to an IR   ``Gaussian fixed point", both   associated to a  scalar model containing, in addition to the two-derivative term 
$z_1 ( \partial \phi \partial \phi)$, a pair of four-derivative terms, namely $z_2 ( \partial^2 \phi \partial^2\phi )$ and $z_3 ( \partial \phi \partial \phi)^2$, 
 with the inessential difference that in \cite{buccio}, instead of directly looking  at the flow of $z_1$ and $z_2$ (as  in our analysis),  
 the flow of the anomalous dimensions associated to these couplings is studied.
 Incidentally, we remark that, in our analysis, we could also have included the term proportional to $z_3$, to account for an even more general model,
 but, because of the result shown in \cite{buccio} on the beta-function of the dimensionless coupling $\widetilde z_3$ : $\beta_{\widetilde z_3}\propto 
 \widetilde z_3$, i.e.  $\widetilde z_3  =0$ is a stationary point for this variable, we preferred to maintain $z_3$ turned off and focus on model 
 \eqref{model}, which already contains a definite phase structure. 
 
In  \cite{buccio}, trajectories connecting  the UV Gaussian point (whose appearance is due to the presence  of the higher derivative terms) 
to the IR Gaussian point  (related to the presence of  the two-derivative term) are in fact pointed out and this is in contrast with 
our findings, where a consistent IR limit requires   $\widetilde z_k<0$. However, this should not come as a surprise, because our result strictly 
depends on the presence of the  sine-Gordon potential and on the flow of the coupling $\widetilde g_k$,  which is instead absent in the former analysis.

Actually, the resulting negative value $\widetilde z_k<0$  in the IR  makes the higher derivative model  \eqref{model} interesting. 
In fact, the contrast of a positive higher-derivative and a negative two-derivative term, can generate a non-uniform modulated ground state 
and the associated  excitations have to be constructed upon this peculiar background. 

Then, it is noticeable that such a model has a  few points in common with scalar models of conformally reduced Euclidean gravity, 
where  an UV fixed point is found \cite{Reuter,Reuter:2008wj,Reuter:2008qx,Machado:2009ph}, 
ensuring the non-perturbative renormalizability of the theory, while the sign in front of the two-derivative term turns out to be negative,
as it occurs in our analysis. It has also been suggested that higher derivative terms could cure the instability produced by the negative 
kinetic term \cite{Bonanno:2013dja}.
In addition, even some evidence that a continuous line of UV fixed points could be present for such a model, when analysed within an
improved approximation scheme \cite{Dietz:2016gzg,Bonanno:2023fij}. Despite the structural differences of the two issues, concerning in particular
the form of the potential,  all these mentioned  features clearly resemble  the findings here discussed for the higher derivative sine-Gordon model
and, therefore,  further investigation about  this point is certainly worthwhile.

\section*{acknoledgements}
GGNA acknowledges support from the project PRIN 2022 -
2022XK5CPX (PE3) SoS-QuBa - ”Solid State Quantum
Batteries: Characterization and Optimization”.
DZ gratefully acknowledges  A. Bonanno for enlightening discussions.

\vfill\eject

\appendix\section{Determination of the RG flow equations }\label{appendixa}

We start with  some useful formulae. Given the four-momenta $p_\mu$, $q_\mu$  appearing in the various n-point functions, 
let us define $\theta$ the angle between $p_\mu$ and $q_\mu$, $p\equiv \sqrt{p^2}, \, q\equiv \sqrt{q^2}$, 
$y=(p+q)^2$ and, by means of the derivation of a composite function we get 
\begin{align}
    \frac{d^2}{dp^2}&=\left(\frac{dy}{dp}\right)^2\frac{d^2}{dy^2}+\frac{d^2y}{dp^2}\frac{d}{dy}\label{der2}\\
    \frac{d^4}{dp^4}&=\left(\frac{dy}{dp}\right)^4\frac{d^4}{dy^4}+6\left(\frac{dy}{dp}\right)^2\frac{d^2y}{dp^2}\frac{d^3}{dy^3}+3\left(\frac{d^2y}{dp^2}\right)^2\frac{d^2}{dy^2}\label{der4}
\end{align}
and, when the derivatives are computed at $p=0$,  we find $y'(p)|_{p=0}=2q\cos(\theta)$ and $y''(p)|_{p=0}=2$.
For   $G_k (y)=( {u_ky^2+z_ky+V''_k+k^4} )^{-1}$ :
\begin{eqnarray}
   && \frac{\partial G_k (y)}{\partial y}=- (2 u_ky+z_k) \, G_k (y)^{2}\nonumber\\
    &&\frac{\partial^2 G_k (y)}{\partial y^2}=2\,(2u_ky+z_k)^2 \; G_k(y)^3 - 2 u_k \; G_k(y)^2
    \nonumber\\
    &&\frac{\partial^3 G_k (y)}{\partial y^3}=12u_k\,(2u_ky+z_k) \; G_k(y)^3- 6(2u_ky+z_k)^3  \; G_k(y)^4\nonumber\\
    &&\frac{\partial^4 G_k (y)}{\partial y^4}=  24u_k^2 \; G_k(y)^3 +24(2u_ky+z_k)^4 \;G_k(y)^5 - 72u_k(2u_ky+z_k)^2 \;G_k(y)^4 \;.
\end{eqnarray}

Finally, in order to avoid the extremely lengthy integration of hypergeometric functions  we resort to  the Schwinger representation:
\begin{equation}\label{proper}
    \frac{1}{A^n}=\int_0^{\infty}ds\frac{s^{n-1}}{(n-1)!}\exp{(-sA)}.
\end{equation}
and we shall also use the following results to integrate out the field $\varphi$:
 \begin{subequations}
\begin{align}
    -\frac{1}{2\pi }\int_{-\pi }^{\pi }d\varphi\cos\varphi\,e^{-s\widetilde{g}_k\cos\varphi}&= I_ 1(\widetilde{g}_k s),\label{bessp}\\
    \frac{1}{2\pi }\int_{0 }^{2\pi} d\varphi\sin^2\varphi\,e^{-s\widetilde{g}_k \cos\varphi}&=\frac{I_ 1(\widetilde{g}_k s)}{\widetilde{g} _ks},\label{bess}\\
    \frac{1}{2\pi}\int_{0 }^{2\pi }d\varphi \sin^2\varphi\cos{\varphi}\,e^{-s\widetilde{g}_k\cos\varphi}&=-\frac{I_ 2(\widetilde{g}_k s)}{\widetilde{g} _ks},\label{bessc}\\
    \frac{1}{2\pi}\int_0^{2\pi}d\varphi\sin ^2\varphi \cos^2\varphi \,e^{-s\widetilde{g}_k\cos\varphi}&=\frac{\widetilde{g}_ksI_
1(\widetilde{g}_ks)-3I_ 2(\widetilde{g}_ks)}{\widetilde{g}_k^2 s^2}.\label{besscc}
\end{align}
\end{subequations} 
where $I_a(x)$ is the modified Bessel function of the first kind \cite{grad_riz}.

\subsection{The equation for {$\widetilde u_k$}}
The flow equation for $\widetilde u_k=u_k$,  is obtained from  \eqref{uk} (note that the integral is written in $d$ dimensions and only at the end 
of the computation we shall specialise it to $d=4$)
and from \eqref{der4} and its structure is 
$P_0[U_1+U_2+U_3]$ where  $P_0$ is given in Eq. \eqref{proj0}
and the three terms $U_1,U_2,U_3$
are respectively related to the second,  the third, and the fourth derivative of $G_k (y)$ with respect to $y$.

After carrying  out the field integration  by making use of  \eqref{bess} and \eqref{bessc}, we obtain for $P_0 U_1$
(here $y=q^2$):
\begin{eqnarray}
\frac{P_0}{2}&&\int\frac{d^dq}{(2\pi)^{d}} \;  \partial_tR_k \; g_k^2 \; \sin^2{\varphi} \; \Big [ 2(2u_k \, q^2+z_k)^2\; G_k^5(q^2)-2u_k\;G_k^4(q^2)) \Big ]=
\nonumber\\
-\frac{4dk^{d-4}\widetilde{g}_k^2}{(4\pi)^{d/2}\Gamma(1+d/2)}&&\int_0^{2\pi}d\varphi\int_0^{\infty}dqq^{d-1}
\sin^2{\varphi}\frac{3u_k^2q^4+3u_k\widetilde{z}_kq^2+\widetilde{z}_k^2-u_k-u_k\widetilde{g}_k\cos{\varphi}}{(u_kq^4+\widetilde{z}_kq^2+\widetilde{g}_k\cos{\varphi}+1)^5}=
\nonumber\\
-\int_0^{\infty}ds \int_{0}^{\infty}dy\, y^{d/2-1}&&
\frac{s^4 \,k^{d-4} \,d \widetilde{g}^2_k \;   e^{-s(1+u_ky^2+\widetilde{z}_ky)}  }{12 (4\pi)^{d/2}\Gamma(1+d/2)} 
\left[\frac{(3u_k^2y^2+3u_k\widetilde{z}_ky+\widetilde{z}_k^2-u_k)I_ 1(\widetilde{g}_ks)}{\widetilde{g}_ks}+\frac{u_kI_2(\widetilde{g}_ks)}{s}\right] \, . \;\;\;\;\;\;\;
\end{eqnarray}
 Then momentum integration cannot be performed for generic dimension $d$ but, by  taking the limit $d\to 4^+$ before performing the proper time integral
 we find (Erf (z) is the error function \cite{grad_riz})
\begin{eqnarray}
P_0 U_1=&&\int_0^{\infty}ds\frac{\widetilde{g}_k e^{-s} s}{1536 \pi ^2 u_k^2}\Biggl \{ I_1(\widetilde{g}_k s) 
\Biggl  [ \sqrt{\pi } \sqrt{s u_k} \,\widetilde{z}_k \,e^{\frac{s \widetilde{z}_k^2}
{4 u_k}} \Biggl( (4 s-6) u_k-s \widetilde{z}_k^2  \; \text{Erf}\left(\frac{\sqrt{s} \widetilde{z}_k}{2 \sqrt{u_k}}\right) +  \nonumber \\
&&8 (s-3) u_k^2-2 u_k \widetilde{z}_k \Bigl(\sqrt{\pi } (2 s-3) \sqrt{s u_k} e^{\frac{s \widetilde{z}_k^2}{4 u_k}}+ s \widetilde{z}_k\Bigr)
 +\sqrt{\pi } s \widetilde{z}_k^3 \sqrt{s u_k} e^{\frac{s \widetilde{z}_k^2}{4 u_k}}\Biggr)\Biggr ]-  \nonumber\\
 &&4 \widetilde{g}_k s u_k I_2(\widetilde{g}_k s) \Biggl(\sqrt{\pi } \sqrt{s} \sqrt{u_k} \widetilde{z}_k 
 e^{\frac{s \widetilde{z}_k^2}{4 u_k}} \text{Erf}\left(\frac{\sqrt{s} \widetilde{z}_k}{2 \sqrt{u_k}}\right) 
-\sqrt{\pi } \widetilde{z}_k \sqrt{s u_k} e^{\frac{s \widetilde{z}_k^2}{4 u_k}}+2 u_k\Biggr)\Biggr \} \nonumber.
\end{eqnarray}

\vskip 5 pt
Analogously, for $P_0 U_2 $ we find 
\begin{eqnarray}
&&\frac{P_0}{4!}\int\frac{d^dq}{(2\pi)^d}\; \partial_t R_k \; g_k^2 \,  \sin^2{\varphi} \,   \cos^2{\theta}\; 48 q^2 
 \Biggl(12 u_k(2u_k q^2+z_k) G_k^5(q^2) - 6 (2u_k q^2+z_k)^3 G_k^6(q^2)\Biggr) = \nonumber\\
&&-\frac{8 k^{d-4} \widetilde{g}_k^2}{(4\pi)^{d/2} \Gamma(1+d/2)} P_0 \int_0^{\infty} dq \, q^{d+1} \sin^2{\varphi} 
 \Biggl[ (12u_k(2u_kq^2+\widetilde{z}_k)(u_kq^4+\widetilde{z}_kq^2+1)-6(2u_kq^2+\widetilde{z}_k)^3) \nonumber\\
&&
+ 12u_k\widetilde{g}_k (2u_kq^2+\widetilde{z}_k)\cos{\varphi} \Biggr] G_k^6(q^2),
\end{eqnarray}
Then, by performing  the field integration through \eqref{bess} and \eqref{bessc}, the momenta integration and by taking the limit  $d\to 4^+$ 
we get:
\begin{eqnarray}
&&P_0U_2=\int_0^{\infty}ds\frac{\widetilde{g}_k e^{-s} s}{{640 \pi ^2 u_k^{3/2}}}
\Biggl(\sqrt{\pi } \sqrt{s} \widetilde{z}_k e^{\frac{s \widetilde{z}_k^2}{4 u_k}} \Biggl(\text{Erf}\left(\frac{1}{2} \widetilde{z}_k \sqrt{\frac{s}{u_k}}\right)-1\Biggr) \times
\nonumber  \\
&&\Biggl(I_1(\widetilde{g}_k s) \Biggl((6-4 s) u_k+s \widetilde{z}_k^2\Biggr)+4 \widetilde{g}_k s u_k I_2(\widetilde{g}_k s)\Biggr) +\nonumber\\ 
&&2 \sqrt{u_k} \Biggl(I_1(\widetilde{g}_k s) \Biggl(s \widetilde{z}_k^2-4 (s-2) u_k\Biggr)+4 \widetilde{g}_k s u_k I_2(\widetilde{g}_k s)\Biggr)\Biggr).
\end{eqnarray}

For the term $P_0U_3$ we have
\begin{eqnarray}
&&\frac{P_0}{4!}\int\frac{d^dq}{(2\pi)^d}\; \partial_t R_k \, g_k^2 \; \sin^2{\varphi} \, \cos^4{\theta}  \, 16 q^4\times \nonumber\\
&&\Biggl(24 (2u_k q^2+z_k)^4 G_k^7(q^2) - 72 u_k (2u_k q^2+z_k)^2 G_k^6(q^2) + 24 u_k^2 G_k^5(q^2)\Biggr) = \nonumber \\
&&- \; \frac{64 k^{d-4} 3 \widetilde{g}_k^2}{4!(d+2) (4\pi)^{d/2} \Gamma(1+d/2)} P_0\int_0^{\infty} dq \, q^{d+3} \Biggl[ 24 (2u_k q^2+\widetilde{z}_k)^4 - \nonumber\\
&&72 u_k (2u_k q^2+\widetilde{z}_k)^2 (u_k q^4 + \widetilde{z}_k q^2 + \widetilde{g}_k \cos{\varphi} + 1)
+ 24 u_k^2 (u_k q^4 + \widetilde{z}_k q^2 + \widetilde{g}_k \cos{\varphi} + 1)^2 \Biggr] G_k^7(q^2).
\end{eqnarray}
The integration over  $\varphi$ is done with the help of  \eqref{bess},\eqref{bessc} and also \eqref{besscc}. 
Finally, by integrating over the momenta e performing the  limit $d\to4^+$ we obtain the third contribution to the flow of $u_k$  
(Erfc (z) is the complementary error function \cite{grad_riz}):
\begin{eqnarray}
&&P_0 U_3 = \int_0^{\infty} ds \frac{g e^{-s} s}{368640 \pi^2 u_k^4}\Biggl \{ I_1(\widetilde{g}_k s) \Biggl [ \sqrt{\pi} \widetilde{z}_k e^{\frac{s \widetilde{z}_k^2}{4 u_k}} 
\text{Erfc}\left(\frac{1}{2} \widetilde{z}_k \sqrt{\frac{s}{u_k}}\right) 
 \Biggl( 8 s^2 \widetilde{z}_k^2 (s u_k)^{3/2} \left(2 (\widetilde{g}_k^2+1) u_k - \widetilde{z}_k^2\right)  +\nonumber \\
&& 2 s (s u_k)^{3/2} \Bigl( 48 (\widetilde{g}_k^2+1) u_k^2 - 64 u_k \widetilde{z}_k^2 + 13 \widetilde{z}_k^4 \Bigr)+ \widetilde{z}_k^6 \sqrt{s^7 u_k} 
+ 1800 \sqrt{s u_k^7} +
\nonumber \\
&& 60 u_k (s u_k)^{3/2} (7 \widetilde{z}_k^2 - 24 u_k) \Biggr)+ 2 u_k \Biggl( - 64 u_k^3 \Bigl( s (\widetilde{g}_k^2 s+s-20) + 30 \Bigr) - \nonumber \\
 &&4 s u_k^2 \widetilde{z}_k^2 \Bigl( 4 s (\widetilde{g}_k^2 s+s-7) + 95 \Bigr)- s^3 \widetilde{z}_k^6 + 8 (s-3) s^2 u_k \widetilde{z}_k^4 \Biggr) \Biggr ] +
 \nonumber \\
&& 8 \widetilde{g}_k s u_k^{3/2} I_2(\widetilde{g}_k s) \Biggl [ \sqrt{\pi} \sqrt{s} \widetilde{z}_k e^{\frac{s \widetilde{z}_k^2}{4 u_k}}
\Bigl( s^2 \widetilde{z}_k^4 - 24 (s-6) u_k^2 + 2 s (5-2 s) u_k \widetilde{z}_k^2 \Bigr) \text{Erfc}\left(\frac{1}{2} \widetilde{z}_k \sqrt{\frac{s}{u_k}}\right) +
\nonumber \\
&& 2 \sqrt{u_k} \Bigl( - s^2 \widetilde{z}_k^4 + 8 (2 s-17) u_k^2 + 4 (s-2) s u_k \widetilde{z}_k^2 \Bigr) \Biggr ] \Biggr\} \; .
\end{eqnarray}
The sum of the three contributions yields the flow equation for $u_k$ :
\begin{equation}\label{flowu}
 \partial_t u_k=\int_0^{\infty}ds\frac{\widetilde g s e^{-s} }{368640 \pi ^2 u_k^4} \left [ c_1I_1(\widetilde{g}_k s)+c_2I_2(\widetilde{g}_k s) \right],
\end{equation}
where 
\begin{align*}
c_1 &= \Biggl( \sqrt{\pi} \widetilde z_k e^{\frac{s \widetilde z_k^2}{4 u_k}} \text{Erfc}\left(\frac{1}{2} \widetilde z_k \sqrt{\frac{s}{u_k}}\right) 
\Biggl( 8 s^2 \widetilde z_k^2 (s u_k)^{3/2} \left(2 ( \widetilde g_k^2+1) u_k - \widetilde z_k^2\right) \\
& \quad + 2 s (s u_k)^{3/2} \Biggl( 48 (\widetilde g_k^2+1) u_k^2 - 64 u_k \widetilde z_k^2 + 13 \widetilde  z_k^4 \Biggr) 
+ \widetilde z_k^6 \sqrt{s^7 u_k} - 216 \sqrt{s u_k^7} \\
& \quad + 12 u_k (s u_k)^{3/2} \left(7 \widetilde z_k^2 - 8 u_k \right) \Biggr) + 2 u_k \Biggl( - 64 u_k^3 \Bigl( (\widetilde g_k^2+1) s^2 + s + 3 \Bigr) \\
& \quad - 4 s u_k^2 \widetilde z_k^2 \Bigl( 4 s \left( \widetilde g_k^2 s + s - 7\right) + 11 \Bigr) - s^3 \widetilde z_k^6 + 8 (s-3) s^2 u_k 
\widetilde z_k^4 \Biggr) \Biggr); \\
\\
c_2 &= 8 \widetilde g_k s u_k^{3/2} \Biggl( \sqrt{\pi} \sqrt{s} \widetilde z_k e^{\frac{s \widetilde z_k^2}{4 u_k}} \Bigl( s^2 \widetilde z_k^4 - 24 (s+1) u_k^2 
+ 2 s (5-2 s) u_k \widetilde z_k^2 \Bigr) \\
& \quad \text{Erfc}\left(\frac{1}{2} \widetilde z_k \sqrt{\frac{s}{u_k}}\right) + 2 \sqrt{u_k} \Bigl( -s^2 \widetilde z_k^4 + 16 (s+2) u_k^2 + 4 (s-2) s u_k  
\widetilde z_k^2 \Bigr) \Biggr).
\end{align*}

\subsection{The equation for {$\widetilde z_k$}}

To calculate the flow equation of $\widetilde z_k$ we start from Eq. \eqref{zk} (note that the integral is written in $d$ dimensions and only at the end 
of the computation we shall specialise it to $d=4$)
and , from Eq.   \eqref{der2}, we notice that its structure in this case is of the form: $P_0[Z_1+Z_2]$, where, as before, $Z_1$ is related to the first derivative,
 while $Z_2$ is related to the second derivative of $G_k(y)$.
By proceeding as for the $u_k$ case, we get
\begin{eqnarray}
 &&\frac{P_ 0}{2!}\int\frac{d^dq}{(2\pi)^d}\;\partial_tR_k\; g_k^2 \; \sin^2{\varphi}\; \Bigg( -(2u_kq^2+z_k)G_k^4(q^2) \Bigg)=
 \nonumber\\
 && \frac{k^{d-2}\, \widetilde{g}_k^2\, 4d }{(4\pi)^{d/2}\Gamma(1+d/2)}
 \int_0^{\infty}ds\frac{s^3}{3!} \; P_ 0\int_0^{\infty}dq q^{d-1}\sin^2{\varphi}\; (2u_kq^2+\widetilde{z}_k) 
 e^{-s(u_kq^4+\widetilde{z}_k q^2+\widetilde{g}_k\cos{\varphi}+1)} =
 \nonumber\\
 &&\frac{k^{d-2}\widetilde{g}^2_k\, 2d }{(4\pi)^{d/2}\Gamma(1+d/2)}\int_0^{\infty}ds\; e^{-s} \frac{s^3}{3!}\int_0^{\infty}dy\,y^{d/2-1}\,(2u_ky+\widetilde{z}_k)
 e^{-s \, (u_ky^2+\widetilde{z}_ky)}\frac{I_ 1(\widetilde{g}_ks)}{\widetilde{g}_ks} \;,
 \end{eqnarray}
 where in the last term we performed the field integration  via Eq. \eqref{bess} and, again,  $y=q^2$.
After performing the momentum  integration and taking the limit $d\to 4^+$,  we find :
\begin{equation}
P_0 Z_1=\int_0^{\infty}ds\left[ -\frac{\widetilde{g}_k k^2 s I_1(\widetilde{g}_k s) e^{\frac{1}{4} s \left(\frac{\widetilde{z}_k^2}{u_k}-4\right)} 
\left(\text{Erf}\left(\frac{1}{2} \widetilde{z}_k \sqrt{\frac{s}{u_k}}\right)-1\right)}{48 \pi ^{3/2} \sqrt{s u_k}}\right].
\label{p0z1}
\end{equation}

We now turn to $P_0Z_2 $ and, after integrating the angular part and making use of  Eqs.  \eqref{bess}, \eqref{bessc}, \eqref{besscc}, to integrate the 
field $\varphi$, we obtain
\begin{eqnarray}
&&\frac{P_ 0}{2!}\int\frac{d^dq}{(2\pi)^d}\partial_tR_kg_k^2\sin^2{\varphi}4q^2\cos^2{\theta}(2(2u_kq^2+z_k)^2G^5(q)-2u_kG^4(q))=
\nonumber\\
&&-\frac{8k^{d-2}\widetilde{g}^2_k}{(4\pi)^{d/2}\Gamma(1+d/2)}\int_0^{\infty}dse^{-s}\frac{s^4}{4!}P_ 0
\int_{0}^{\infty}q^{d+1}e^{-s(u_kq^4+\widetilde{z}_kq^2+\widetilde{g}_k\cos{\varphi})} 
\times
\nonumber\\
&&2 \sin^2{\varphi}(3u_k^2q^4+3u_k\widetilde{z}_kq^2+\widetilde{z}^2_k-u_k-u_k\widetilde{g}_k\cos{\varphi}) =
\nonumber\\
 && -\frac{8k^{d-2}\widetilde{g}^2_k}{(4\pi)^{d/2}\Gamma(1+d/2)}\int_0^{\infty}e^{-s}\frac{s^4}{4!}\int_0^{\infty}dyy^{d/2}e^{-s(u_ky^2+\widetilde{z}_ky)}
 \times \nonumber\\&& 
\left((3u_k^2y^2+3u_k\widetilde{z}_ky+\widetilde{z}^2_k-u_k)\frac{I_1(\widetilde{g}_ks)}{\widetilde{g}_ks}+u_k\frac{I_ 2(\widetilde{g}_ks)}{s}\right) \; .
\end{eqnarray}

Performing the momentum integration and  taking the  limit $d\to 4^+$ yield
\begin{eqnarray}
&&
P_0Z_2=-\int_0^{\infty}ds\;\frac{\widetilde{g}_k k^2\sqrt{s}e^{-s}}{3072 \pi ^2 u_k^{5/2}\widetilde{z}_k}\Biggl\{ I_1(\widetilde{g}_k s) \Biggl[ \widetilde{z}_k\Biggl(-2 \widetilde{z}_k^3 \sqrt{s^3 u_k}+\sqrt{\pi } e^{\frac{s \widetilde{z}_k^2}{4 u_k}} \Bigl [ s^2 \widetilde{z}_k^4 
\nonumber\\&&
+4 (9-2 s) u_k^2-4 (s-2) s u_k \widetilde{z}_k^2\Bigr ] -12 \widetilde{z}_k \sqrt{s u_k^3}+8 \widetilde{z}_k (s u_k)^{3/2}\Biggr) 
\nonumber\\&&
-\sqrt{\pi } \widetilde{z}_k e^{\frac{s \widetilde{z}_k^2}{4 u_k}} \Biggl(s^2 \widetilde{z}_k^4
+4 (9-2 s) u_k^2-4 (s-2) s u_k \widetilde{z}_k^2\Biggr)\text{Erf}\left(\frac{1}{2} \sqrt{\frac{s \widetilde{z}_k^2}{u_k}}\right) \Biggr ] \Biggr\} 
\nonumber\\&&
+4 \widetilde{g}_k s u_k \sqrt{\widetilde{z}_k^2} I_2(\widetilde{g}_k s) \Biggl [ \sqrt{\pi } e^{\frac{s \widetilde{z}_k^2}{4 u_k}} 
\left(s \widetilde{z}_k^2+2 u_k\right) \text{Erfc}\left(\frac{1}{2} \widetilde{z}_k \sqrt{\frac{s}{u_k}}\right)-2 \widetilde{z}_k \sqrt{s u_k}\Biggr)\Biggr].
\label{p0z2}
\end{eqnarray}

Finally, the sum of Eq. \eqref{p0z1}  and  Eq. \eqref{p0z2}  provide the flow equation  for $\widetilde{z}_k$
\begin{equation}\label{flowz}
 \partial_t \widetilde z_k=2\widetilde{z}_k+\int_0^{\infty}ds\frac{\widetilde{g}_k e^{-s} \left(c_3 I_ 1(\widetilde{g}_k s)-c_4 I_ 2(\widetilde{g}_k s)\right)}{3072\pi^2},   
\end{equation}
where we used:
\begin{align*}
c_3&=\frac{\sqrt{s}}{u_k^{5/2}}\Biggl [ \sqrt{\pi } e^{\frac{s \widetilde{z}_k^2}{4 u_k}} \Big ( -s^2 \widetilde{z}_k^4+4 (2s+7) u_k^2+4 (s-2) s u_k \widetilde{z}_k^2 \Big) 
\;\text{Erfc}\left(\frac{1}{2} \widetilde{z}_k
\sqrt{\frac{s}{u_k}}\right)\\
&+2 \widetilde{z}_k \Big (\widetilde{z}_k^2 \sqrt{s^3 u_k}+6 \sqrt{su_k^3}-4 (s u_k)^{3/2}\Big) \Biggr];\\
c_4&=4 \widetilde{g}_k \left(\frac{s}{u_k}\right)^{3/2}\left [ \sqrt{\pi } e^{\frac{s \widetilde{z}_k^2}{4 u_k}} \left(s\widetilde{z}_k^2+2 u_k \right) \text{Erfc}\left(\frac{1}{2} \widetilde{z}_k \sqrt{\frac{s}{u_k}}\right)-2 \widetilde{z}_k \sqrt{s u_k}\right]. 
\end{align*}

\subsection{The equation for $\widetilde g_k$}

To calculate the flow equation of $\widetilde g_k$ we start from Eq. \eqref{Vflow} (note that the integral is written in $d$ dimensions and only 
at the end of the computation we shall specialise it to $d=4$)
and use the projector $P_1$ defined in  Eq. \eqref{proj1}.
\begin{eqnarray}
&&P_ 1\partial_tV_k=\frac{P_1}{2}\int\frac{d^dq}{(2\pi)^d}\frac{\partial_tR_k}{u_kq^4+z_kq^2+g_k\cos{\varphi}+k^4}=
\nonumber\\
&&-\frac{k^d\; 2d}{(4\pi)^{d/2}\Gamma(1+d/2)}
\int_{0}^{\infty}ds\,e^{-s}\int_0^{\infty}dqq^{d-1}\,e^{-s(u_kq^4+\widetilde{z}_kq^2)}\left(-\frac{1}{\pi}\int_{-\pi}^{\pi}d\varphi\cos\varphi\,
e^{-s\widetilde{g}_k\cos{\varphi}}\right) \, .
\end{eqnarray}
We now use Eq. \eqref{bessp} to integrate over the field, thus obtaining the following equation in terms of the variables  $u_k$, $\widetilde z_k$, $\widetilde g_k$:
\begin{equation}
\partial_t\widetilde{g}_k=4\widetilde{g}_k -\frac{2dk^{d-4}}{(4\pi)^{d/2}\Gamma(1+d/2)}
\int_0^{\infty}\
ds\exp{(-s)}I_ 1(\widetilde{g}_k \
s)\int_0^{\infty}dq^2(q^2)^{d/2-1}\exp{(-s(u_kq^4+\widetilde{z}_kq^2))}.
\end{equation}

After performing the momentum integral and taking the limit $d\to 4^+$ we find

\begin{equation}\label{flowg}
\partial_t\widetilde{g}_k=
4\widetilde{g}_k-\frac{1-\sqrt{1-\widetilde{g}_k^2}}{8\pi^2u_k\widetilde{g}_k}+\frac{\widetilde{z}_k}{(4\pi)^{2}}
\int_0^{\infty}ds \, e^{-s} \,\, \frac{I_ 1(\widetilde{g}_k s) }{s u_k}
\left [ {\sqrt{\pi } \sqrt{\frac{s}{u_k}} e^{\frac{s \widetilde{z}_k^2}{4 u_k}}\; \text{Erfc}\left(\frac{\sqrt{s} \widetilde{z}_k}{2\sqrt{u_k}}\right)}\right] \;.
\end{equation}

\section{RG flow from perturbative expansion\label{appendixb}}

\subsection{The four-dimensional  case}

In this Appendix  we make use of the mapping,  discussed in \cite{defenu},  between the model in Eq. \eqref{model} with $z=0$ and 
a generalisation of the  Coulomb gas in four dimensions, with the aim of deriving in an alternative way the RG equations for the couplings
and, in particular, to show that the negative sign in Eq. \eqref{flowuredu} is actually the opposite of the sign in the two-dimensional case, as it is 
strictly related to the number of spatial dimensions of the problem.
We start from the Hamiltonian density
\begin{equation}
H=\frac{\cal K}{2} \int d^4 {\bf r} \; \Big [ \Delta \theta ({\bf r})  \Delta \theta ({\bf r}) + {\cal T} \, \nabla_i \theta ({\bf r})  \nabla_i \theta ({\bf r})    \Big ]
\label{accaeff3}
\end{equation}
that, in addition to the Laplacian  of the field $\theta$ squared,
contains an additional lower derivative term, proportional to the new coupling 
${\cal T}$ which will be used as a counterterm of higher order corrections appearing in our computation.
As explained in  Appendix B of \cite{defenu}  the Green's function of the square Laplacian $\Delta^2$ in $d=4$ is 
\begin{equation}
{\cal G} ({\bf r}-  {\bf r'} ) \; = \; \int \frac{d^4{\bf r''} }{(2\pi)^2}  \frac{1}{({\bf r} - {\bf r''})^{2}} \frac{1}{({\bf r''} - {\bf r'})^{2}} \;=\;
- \; \frac { 1}{2}\; {\rm  ln} \frac{ \left | {\bf r} - {\bf r'} \right |}{R}
\label{configapp} 
\end{equation} 
that corresponds to the response at the point  ${\bf r}$
of a charge located in  ${\bf r'}$  while $R$ indicates the size of the system. The right-hand side of 
(\ref {configapp}) comes from the explicit resolution of the integral.

The form of  ${\cal G} ({\bf r}-  {\bf r'} )$ in Eq. (\ref {configapp}) comes from the general property of the Laplacian in $d=4$ \cite{evans}
\begin{equation}
\Delta  \;\; {({\bf r} - {\bf r'})^{-2}} = - \;(2\pi)^2 \delta^4 ({\bf r} - {\bf r'})
\label{lapsol}
\end{equation}
which gives 
\begin{equation}
\Delta^2\, {\cal G} ({\bf r}-  {\bf r'} )  = (2\pi)^2 \delta^4 ({\bf r} - {\bf r'}) \;,
\label{lapsol2}
\end{equation}
or equivalently 
\begin{eqnarray}
\Delta  \;\; {\rm  ln}  \left | {\bf r} - {\bf r'} \right | &=& 2 \;\; ({\bf r} - {\bf r'})^{-2} \;\;\;\;\;\;\;\;\;,
\label{lapsquare21} \\
\Delta^2\, {\rm  ln}  \left | {\bf r} - {\bf r'} \right |  &=&  -8 \,\pi^2 \delta^4 ({\bf r} - {\bf r'}) \;\;\;.
\label{lapsquare22}
\end{eqnarray}

 Then from Eqs. (\ref{accaeff3},\ref{configapp}, \ref{lapsol2}) one easily finds the following structure of 
 the interaction hamiltonian density associated to a charge-anticharge  pair, respectively located in ${\bf r}$ and ${\bf r'}$ :
\begin{equation}
H^{(2)}= H_{core}+
\frac{\cal K}{2} \, (2\pi)^2  \; {\rm  ln}  \left | {\bf r} - {\bf r'} \right | 
\;+\;\frac{\cal K \; T }{2} \int d^4{\bf r''} \;\frac{{\rm  ln}  \left | {\bf r''} - {\bf r'} \right | }{({\bf r} - {\bf r''} )^2} \;,
\label{hamil2ch} 
\end{equation} 
where $H_{core}$ includes all self-interaction effects of each charge of the pair, which typically have the same form of the 
two terms $O({\cal K })$ and $O({\cal K T})$  in Eq. \eqref{hamil2ch}  when the limit ${\bf r} \to {\bf r'}$ is taken with a suitable
short distance regulator.

\hskip 5 pt
In our computation of  higher order corrections, we will not include the effects of $O({\cal K T})$ term of 
Eq. (\ref{hamil2ch})  
but nevertheless we retain it at the lowest order and treat it as a counterterm to cancel other terms  generated in the procedure. 
In perturbation theory, this is justified if we assume ${\cal K }$ and ${\cal T}$ as small couplings of the same order, 
so that  the $O({\cal K T})$ term is naturally higher order.

Then,  in perturbation theory,  we compute the  charge-anticharge effective interaction, 
that includes dressing effects of their interaction with  further charge-anticharge pairs. In this way, 
the system is kept globally neutral. 
Our  procedure  follows the demonstration developed in \cite{Jose1977}
for the BKT transition in $d=2$,  by rearranging it to our specific problem in $d=4$ and, in particular 
we write down the partition function associated to the a system of interacting charge-anticharge pairs:
\begin{equation}
\label{partition}
{\cal Z}_{_{V}}=\sum_{N=0}^{\infty} \; \frac{y_0^{2N}}{(N!)^2} \; \prod_{i'=1}^{2N}\int \, d^4 x_{i'} \; e^{-H^{(2)}_{I}} \; ,
\end{equation}
where the sum is taken over $N$ charge-anticharge pairs, and the interaction Hamiltonian 
 $H^{(2)}_{I}$   of a pair  is read from Eq. (\ref{hamil2ch}):
 \begin{equation}
H^{(2)}_{I} \left ( |{\bf r} - {\bf r'} |\right ) =H^{(2)} - H_{core} \; ,
\label{accai}
\end{equation}
whereas the effects of $H_{core}$ are collected in the fugacity $y_0=e^{-H_{core}}$, 
associated to the single charge self-energy.
All lengths expressed in units of the lattice spacing $a$.

The effective interaction $H^{(2)}_{eff}$  of a charge-anticharge pair, located at  ${\bf r}$ and ${\bf r'}$ 
is given by
\begin{equation}
e^{-H^{(2)}_{eff} \left ( |{\bf r} - {\bf r'} |\right )  }  = \langle e^{-H^{(2)}_{I} \left ( |{\bf r} - {\bf r'} |\right )   }  \rangle \; ,
\label{accaeff}
\end{equation}
where the average is computed by means of the partition function in Eq. (\ref{partition}), 
perturbatively expanded in powers of the fugacity $y_0$.

We compute the first correction of order $y_0^2$ that corresponds to the effect on the
external charges  located at ${\bf r}$ and ${\bf r'}$, of two  additional internal charges located at ${\bf s}$ and ${\bf s'}$  
and eventually we shall integrate over ${\bf s}$ and ${\bf s'}$, in order to sum over 
all possible spatial configurations of these additional charges.
It is assumed that  the primed coordinates have negative charges, while the unprimed are positive.
To order $O(y_0^2)$ :
\begin{eqnarray}
&&e^{-H^{(2)}_{eff}\left ( |{\bf r} - {\bf r'} |\right ) }  =  e^{-H^{(2)}_{I}  \left ( |{\bf r} - {\bf r'} |\right )   }  \; \frac
{\left\{ 1+ y_0^2 \int d^4 {\bf s} \int d^4 {\bf s'}  \; e^{-2\pi^2 {\cal K } \; {\rm ln}   \left | {\bf s} - {\bf s'} \right | + 
2\pi^2 {\cal K } \, D(  {\bf r}, {\bf r'}, {\bf s}, {\bf s'})}+ O(y_0^4) \;\right\} }
{\left\{ 1+ y_0^2 \int d^4 {\bf s} \int d^4 {\bf s'} \; e^{-2\pi^2 {\cal K } \; {\rm ln}  \left | {\bf s} - {\bf s'} \right | } 
+ O(y_0^4) \;\right\} }  = \nonumber\\
&&
e^{-H^{(2)}_{I} \left ( |{\bf r} - {\bf r'} |\right )}  \;
\left\{ 1+ y_0^2 \int d^4 {\bf s} \int d^4 {\bf s'}  \; e^{-2\pi^2 {\cal K } \; {\rm ln} \left | {\bf s} - {\bf s'} \right |  } 
\left [ e^{ 2\pi^2 {\cal K } \, D(  {\bf r}, {\bf r'}, {\bf s}, {\bf s'})}  -1 \right ] + O(y_0^4) \;\right\} 
\label{eeffective}
  \end{eqnarray}
where the denominator provides the normalization factor and, as stated above, in the $O(y_0^2)$ correction 
in curly brackets, we discarded the $O({\cal K T})$  part of $H^{(2)}_{I}$ and retained the $O({\cal K})$ term only.

\hskip 5 pt The interaction among the external and internal charges is contained in the factor \break
$
D(  {\bf r}, {\bf r'}, {\bf s}, {\bf s'}  )= 
{\rm ln}   \left | {\bf r} - {\bf s} \right | -
{\rm ln}  \left | {\bf r} - {\bf s'} \right | -
{\rm ln}  \left | {\bf r'} - {\bf s} \right | +
{\rm ln}  \left | {\bf r'} - {\bf s'} \right | 
$,
and the signs reflect the convention adopted for the primed and unprimed coordinates.
It is convenient to change integration variables in Eq. (\ref{eeffective}), from ${\bf s}, {\bf s'}$ to  
$\;\;\;
{\bf X}=({\bf s} +  {\bf s'} )/{2}
\;\; ,\;\;
{\bf x}=({\bf s} - {\bf s'} )
$
and expand the term in square brackets in the second line of Eq. (\ref{eeffective}) in powers of ${\bf x}$, as the factor  
in front of the square brackets, $e^{-2\pi^2 {\cal K } \; {\rm ln} \left | {\bf x} \right | }$,  disfavours largely 
separated pairs with $|{\bf x}| \gg 1$.

It is easy to check that $O(x^{2n})$ are exactly cancelled out in the expansion (note that $\nabla_i$ indicates 
the derivative with respect to the $i$-th component of the full argument of the related logarithm)
\begin{eqnarray}&&
D(  {\bf r}, {\bf r'}, {\bf X}, {\bf x}  ) =
{\rm ln}    \left | {\bf r}  - {\bf X}  -  \frac{\bf x}{2} \right | -
{\rm ln}    \left | {\bf r}  - {\bf X}  + \frac{\bf x}{2} \right | -
{\rm ln}    \left | {\bf r'}  - {\bf X}  -  \frac{\bf x}{2} \right | +
{\rm ln}    \left | {\bf r'}  - {\bf X}  + \frac{\bf x}{2} \right | \simeq \nonumber\\
&&
x_i \nabla_i {\rm ln}  \left | {\bf r'} - {\bf X} \right | 
\;-\; x_i \nabla_i {\rm ln} \left | {\bf r} - {\bf X} \right | 
\;+\; \frac{x_i x_j x_k }{24} \nabla^3_{i,j,k}   {\rm ln}  \left | {\bf r'} - {\bf X} \right | 
\;-\; \frac{x_i x_j x_k }{24} \nabla^3_{i,j,k}  {\rm ln}  \left | {\bf r} - {\bf X} \right |   +O(x^5)\, . \;\;\;\;\;\;
\label{dexpand}
\end{eqnarray}
We can neglect $O(x^5)$  as they produce derivatives of order larger than four and are unnecessary to our purpose.
Therefore, due to the presence of only  odd powers  of $x$ in  
Eq. (\ref{dexpand}), we need to keep even powers of $D(  {\bf r}, {\bf r'}, {\bf X}, {\bf x})$ in Eq. (\ref{eeffective}),
as the odd powers vanish due to the  $d^4 {\bf x}$ integration:
\begin{equation}
\int d^4 {\bf X} \int d^4 {\bf x }  \; e^{-2\pi^2 {\cal K } \; {\rm ln} \left | {\bf x} \right | }
\left [ e^{ 2\pi^2 {\cal K } \, D}  -1 \right ] =
\int d^4 {\bf X} \int d^4 {\bf x }  \; e^{-2\pi^2 {\cal K } \; {\rm ln} \left | {\bf x} \right | }
\left [ \sum^{\infty}_{m=1} \frac{(2\pi^2 {\cal K } \, D)^{2m }}{(2m)!} 
\right ] \; .
\label{svilup1}
\end{equation}
After neglecting $O({\cal K}^4)$ and higher orders  terms in  ${\cal K}$, and neglecting as well terms with 
order of derivatives larger than four,
we are left  with the  following $O(x^4)$ and  $O(x^2)$ contributions from the $O({\cal K}^2)$ term:
\begin{eqnarray}
&&\left [ e^{ 2\pi^2 {\cal K} \, D}  -1 \right ] \simeq  2\pi^4 {\cal K}^2 
\, \Bigg [ \Big ( x_i \nabla_i {\rm ln}  \left | {\bf r'} - {\bf X} \right |  \;-\; x_i \nabla_i {\rm ln} \left | {\bf r} - {\bf X} \right |   \Big)^2  + \nonumber\\
&&2 \Big ( x_l \nabla_l {\rm ln}  \left | {\bf r'} - {\bf X} \right |  \;-\; x_l \nabla_l {\rm ln} \left | {\bf r} - {\bf X} \right |   \Big) 
\; \Big (  \frac{x_i x_j x_k }{24} \nabla^3_{i,j,k}   {\rm ln}  \left | {\bf r'} - {\bf X} \right | 
\;-\; \frac{x_i x_j x_k }{24} \nabla^3_{i,j,k}  {\rm ln}  \left | {\bf r} - {\bf X} \right |   \Big ) \Bigg ] \; .
\label{sviluppol}
\end{eqnarray}

These terms can be integrated by parts and, after symmetrization of  the integrands, we get
\begin{eqnarray}
&& e^{-H_{eff} \left ( |{\bf r} - {\bf r'} |\right )    }  = 
e^{-H^{(2)}_{I} \left ( |{\bf r} - {\bf r'} |\right )   }  \; \Biggl\{ 1+ y_0^2 \int d^4 {\bf X} \int d^4 {\bf x}  \; e^{-2\pi^2 {\cal K} \; {\rm ln} 
\left | {\bf x} \right |  }  \cdot
\nonumber\\
&&
\Biggl [  \frac{ \pi^4 {\cal K}^2 x^2}{2}
\,  \Big ( 2 \; {\rm ln}  \left | {\bf r'} - {\bf X} \right | \; \Delta  {\rm ln} \left | {\bf r} - {\bf X} \right |  
\; -\;  {\rm ln}  \left | {\bf r'} - {\bf X} \right | \; \Delta  {\rm ln} \left | {\bf r'} - {\bf X} \right |
\; -\; {\rm ln}  \left | {\bf r} - {\bf X} \right | \Delta  {\rm ln} \left | {\bf r} - {\bf X} \right | \Big) \; +
 \nonumber\\
 &&
\frac{ \pi^4 {\cal K}^2 x^4}{48}
\Big ( 2 \; {\rm ln}  \left | {\bf r'} - {\bf X} \right | \; \Delta^2  {\rm ln} \left | {\bf r} - {\bf X} \right |  
\; -\;  {\rm ln}  \left | {\bf r'} - {\bf X} \right | \; \Delta^2  {\rm ln} \left | {\bf r'} - {\bf X} \right |
\; -\; {\rm ln}  \left | {\bf r} - {\bf X} \right | \Delta^2  {\rm ln} \left | {\bf r} - {\bf X} \right | \Big)
\Biggr ] \Biggr\}  \; . \;\;\;\;\;\;
\label{develop}
  \end{eqnarray}
The $\Delta$ and $\Delta^2$ applied to the logarithms can be solved with the help of  the relations in (\ref{lapsquare21}) and
(\ref{lapsquare22}).
The $\Delta^2$ generates a $\delta$-function that simplifies the computation of the $d^4{\bf X}$ integral
while the $\Delta$  produces a term with  the same structure of the   $O({\cal K T })$ term in  (\ref{hamil2ch}),  namely 
\begin{equation}
e^{-H_{eff}}  = 
e^{-H^{(2)}_{I}}  \; \Biggl\{ 1+ y_0^2   \pi^4 {\cal K}^2  \Bigg [ 2 \, J_2 \;
\int d^4 {\bf X} \, \Bigg ( \frac{ {\rm ln}  \left | {\bf r'} - {\bf X} \right | }{({\bf r} - {\bf X})^2 }  - \frac{ {\rm ln}  \left | {\bf X} \right | }{({\bf X})^2 }\Bigg )
- \frac{J_4}{24} \; (8\pi^2)\; {\rm ln}  \left | {\bf r'} - {\bf r} \right |  \Bigg ]  \Bigg \}  \; ,
\label{develop2}
\end{equation}
where  we  defined 
$
J_n =\int d^4 {\bf x} \; x^n  \; e^{-2\pi^2 {\cal K } \; {\rm ln} \left | {\bf x} \right | } \; .
$
To order $O(y^2_0)$, the content of the curly brackets in  Eq.~(\ref{develop2})  is equivalent  to a  resummed exponential form 
and this yields  the correction to $H^{(2)}_{I}$. We observe the correspondence
of  the the various contributions in square brackets of  (\ref{develop2}) with those of  $H^{(2)}_{I}$  in  Eq. (\ref{hamil2ch}).
In particular, we notice  that the integral  independent of ${\bf r}$ and ${\bf r'}$ in Eq. (\ref{develop2})  corresponds to 
the self-energy contribution  that has to be reabsorbed  in $y^2_0$, 
while  the contribution proportional to the integral in $d^4{\bf X}$ in (\ref{develop2})  
gets summed to the analogous term from Eq. (\ref{hamil2ch}), yielding
\begin{equation}
\Bigg ( -\;\frac{\cal K \; T }{2} +  2 \, J_2 \, y_0^2  \, \pi^4\, {\cal K}^2 \Bigg )\; \int d^4{\bf X} \;\;\;\frac{  {\rm  ln}  |{\bf r'} - {\bf X}|  }{({\bf r} - {\bf X} )^2} \;.
\label{develop4}
\end{equation}
Clearly the whole contribution in Eq. (\ref{develop4})  vanishes  for the particular choice  of  ${\cal T}$ 
that sets to zero  the sum of the two terms in brackets.
This choice requires ${\cal T}\propto {\cal K}$, 
which justifies why we discarded  $O( {\cal K} {\cal T} )$ terms in 
Eq. (\ref{hamil2ch}) when computing higher order corrections. 
In this case, the interaction hamiltonian is proportional to ${\rm  ln}  |{\bf r'} - {\bf r}|$ only and we get
\begin{equation}
e^{-H^{(2)}_{eff}}  =   e^{- (2\pi^2) \, {\cal K} \; {\rm  ln}  \left | {\bf r} - {\bf r'} \right | }  
\;  e^{  -  \frac{{\cal K}^2}{3} \pi^6 J_4 y_0^2 \;\; {\rm ln}  \left | {\bf r} - {\bf r'} \right | } 
\label{eeffective2}
  \end{equation}
that, with the suitable  parametrization of the hamiltonian density $H^{(2)}_{eff} $ in terms 
of the effective coupling ${\cal K}_{eff}$, 
$H^{(2)}_{eff} = (2\pi^2) \,{\cal K}_{eff} \; {\rm  ln}  \left | {\bf r} - {\bf r'} \right |  $, 
produces the relation

\begin{equation}
{\cal K}_{eff}  =  {\cal K} +  \frac{{\cal K}^2 }{6}\,\pi^4 \,y_0^2 \;J_{4} ={\cal K} +  \frac{{\cal K}^2 }{3}\,\pi^6 \,y_0^2\;
\int_1^\infty  d {x } \;  x^{7- 2\pi^2 {\cal K }} \; .
\label{fina}
\end{equation}
The lower limit of integration in Eq.  (\ref{fina}) is 
set by the minimum distance  allowed between charges, which is equal to the lattice spacing
or $x=1$ as all lengths are expressed in units of the lattice spacing.

By inverting Eq. (\ref{fina}) and neglecting higher orders in ${\cal K}$, we obtain
\begin{equation}
{\cal K}^{-1}_{eff}  ={\cal K}^{-1} -   \frac{\pi^6  y_0^2}{3} \;
\int_1^\infty  d {x } \;  x^{7- 2\pi^2 {\cal K }}  \; .
 \label{newfina}
 \end{equation}

An RG transformation on the couplings ${\cal K}^{-1}$ and $y_0$ is arranged
by requiring the invariance of ${\cal K}_{eff}$ in Eq. (\ref{newfina}) under a rescaling of the lattice spacing.
This is achieved by splitting the integral in (\ref{newfina}) into one integral from 1 to $b>1$ and a second integral
from $b$ to $\infty$ :
\begin{equation}
{\cal K}^{-1}_{eff}  ={\cal K}^{-1} -   
\frac{\pi^6  y_0^2}{3} \;\left ( \int_1^b d {x } \;  x^{7- 2\pi^2 {\cal K }}  + \int_b^\infty  d {x } \;  x^{7- 2\pi^2 {\cal K }} \right ) \;.\label{redefinition}
 \end{equation}
 
As a second step, we define a new $b$-dependent coupling $\widetilde{\cal  K}$ 
\begin{equation}
\widetilde{\cal K}^{-1}  = {\cal K}^{-1}  - \frac{\pi^6  y_0^2}{3} \; \int_1^b  d {x } \;  x^{7- 2\pi^2 {\cal K }} 
\label{ktilde}
\end{equation}
and, after rescaling $x \to b \widehat x$ in the second integral of Eq.  (\ref{newfina}) so that the integral in $d \widehat x$
goes  from 1 to $\infty$, we define the coupling $\widetilde y_0$ containing all the  $b$-dependence
that appears after the $x$ rescaling:
\begin{equation}
\widetilde y_0^2  =  y_0^2 \; b^{8- 2\pi^2 {\cal K }} \;.
\label{ytilde}
\end{equation}

These definitions allow us to rewrite the right  hand side of Eq. (\ref{newfina})  in the same form, but with 
${\cal K}$ and $y_0$  replaced by $\widetilde{\cal  K}$ and $\widetilde y_0$, while leaving 
${\cal K}_{eff}$ unchanged in the left-hand side 
(more precisely, this procedure does not yield  the replacement of ${\cal K}$ in the exponent of $x$ in Eq. (\ref{newfina}),
although the latter substitution is admissible if we neglect higher orders of ${\cal K}$ in the exponent).
Therefore Eqs. (\ref{ktilde}) and (\ref{ytilde}) are the sought-after  RG transormations that can be easily put in  
differential form:
\begin{equation}
b \frac{\partial \widetilde{\cal K}^{-1} }{\partial b} = - \frac{\pi^6}{3} \;  \widetilde y_0^2   \;\;\;  \;\;\; ; \;\;\;\;\;\; \;\;\;  \;\;\; 
b \frac{\partial \widetilde y_0 }{\partial b} = \widetilde y_0 \;\left ( 4- \frac{\pi^2}{  \widetilde{\cal K}^{-1} }\right ) \; .
\label{rgdiff}
\end{equation}
The flow  equations in Eq. (\ref{rgdiff}) presents the same structure of the flow observed in Eqs . \eqref{flowuredu} and \eqref{flowgredu},
at least by neglecting higher order in the couplings,
provided one suitably relates  $\widetilde{\cal  K}^{-1} $ and $\widetilde y_0$  respectively to $u_k$ and $\widetilde g_k$.
In particular the negative sign in the right hand side of the flow equation of  $\widetilde{\cal  K}^{-1}$ is 
confirmed.

\subsection{The general d-dimensional  case}

The same approach illustrated so far in this Appendix, 
when applied the BKT transition in $d=2$, as it was originally formulated in \cite{Jose1977},
turns out to be even simpler, as the interaction hamiltonian density of Eq. (\ref{hamil2ch}), in this case reduces to 
\begin{equation}
 H_{2d}^{(2)}= H_{core}+ 2\pi\,K  \; {\rm  ln}  \left | {\bf r} - {\bf r'} \right | 
\label{hamil2ch2d} 
\end{equation} 
and the Green's function of the Laplacian in $d=2$  is 
\begin{equation}
\Delta \;
{\rm ln}  \left | {\bf r} - {\bf X} \right | = 2\pi \; \delta^2 ({\bf r} - {\bf X})\; .
\label{laplace}
\end{equation}
By implementing the above changes in the procedure followed 
in the previous section, and in particular by retaining only the 
2-derivative contribution in the computation of the corrections,
which is relevant in $d=2$,
we arrive at the flow equations
\begin{equation}
b \frac{\partial \widetilde{ K}^{-1} }{\partial b} =  {4\pi^3}\;  \widetilde y_0^2   \;\;\;  \;\;\; ; \;\;\;\;\;\; \;\;\;  \;\;\; 
b \frac{\partial \widetilde y_0 }{\partial b} = \widetilde y_0 \;\left ( 2- \frac{\pi}{  \widetilde{ K}^{-1} }\right )  \; ,
\label{rgdiff2d}
\end{equation}
which display a positive right-hand side  in the flow equation of  $\widetilde{ K}^{-1}$. This switching of sign between 
Eqs.~(\ref{rgdiff}) and  (\ref{rgdiff2d}) can be traced back to the change of sign 
between Eqs. (\ref{lapsquare22}) and  (\ref{laplace}) that are used respectively in $d=4$ and in $d=2$.

Therefore, the essential difference observable on the flow of the couplings when the number of dimensions of the system  is 
changed, is essentially due to the different sign coming from the application of the suitable power of the Laplacian to 
the logarithmic function.

This argument can be generalized to generic even dimension $d$. In fact, in this case the approach 
discussed so far in $d=4$ and in $d=2$  is recovered by considering the following hamiltonian density 
\begin{equation}
H=\int d^d x \; \; \nabla_i^{d/2}\theta \; \;\; \nabla_i^{d/2}\theta
\label{ham}
\end{equation}
and the essential point is that after integrating by parts 
we end up with the computation of $\Delta^{d/2} \; {\rm  ln}  |{\bf X} - {\bf r}|$,
which provides different signs depending on $d$,
\begin{equation}
\Delta^{\frac{d}{2}} \; {\rm  ln}  |{\bf X} - {\bf r}| = (-1)^{\frac{d}{2}+1} \; \alpha^2_d \; \delta^d  (|{\bf X} - {\bf r}|) \; ,
\label{rfond}
\end{equation}
where $\alpha^2_d$ is a positive constant.  Then, 
the sign of the right hand side of Eq. (\ref{rfond}) is directly related to the sign 
of  the coefficient already determined in the right hand side of the first flow equation 
in Eq. (\ref{rgdiff}) and in Eq. (\ref{rgdiff2d}).
Accordingly, in general  the sign of the  corresponding flow in $d$ dimensions changes as $(-1)^{\frac{d}{2}+1}$.

The origin of the sign  in  Eq. (\ref{rfond})
can be traced back to the specific form of the Laplacian Green's function
for $d>2$ (the $d=2$ case is  in Eq.(\ref{laplace})  ), reported  in \cite{evans} (here $\beta^2_d$ is a positive constant),
\begin{equation}
\Delta \; \frac{1} {|{\bf X} - {\bf r}|^{d-2} }\; = \; - \; \beta^2_d \; \delta^d  (|{\bf X} - {\bf r}|) \; .
\label{lapld}
\end{equation}
Indeed, if we now rename the Green's function  
\begin{equation}
{\cal S}_d ( {\bf s_1}, {\bf s_2} )= \frac{1} {|{\bf s_1} - {\bf s_2}|^{d-2} } 
\label{defi}
\end{equation}
and define the following convolution
\begin{equation}
{\cal C}_d ( {\bf X}, {\bf r} )= \Biggl ( \prod_{k=1}^{\frac{d}{2}-1} \int \; d^d{s_k} \Biggr ) \;\; {\cal S}_d ( {\bf X}, {\bf s_1} ) \;\;   \Biggl ( \prod_{j=1}^{\frac{d}{2}-2 } \; 
{\cal S}_d ( {\bf s_j}, \,{\bf s_{j+1}} ) \,  \Biggl ) 
\;\;{\cal S}_d ( {\bf s_{\frac{d}{2}-1}}, {\bf r} ) \; ,
\label{defi2}
\end{equation}
with the help of Eqs. (\ref{lapld}) and (\ref{defi}), it is  easy  to verify that:
\begin{equation}
\Delta^{\frac{d}{2}} \; {\cal C}_d ( {\bf X}, {\bf r} )  = (-\beta^2_d )^{\frac{d}{2}} \; \delta^d  (|{\bf X} - {\bf r}|)
\label{goal1}
\end{equation}
and also, by direct inspection, to check the following relation ($\gamma^2_d$ is a positive constant),
\begin{equation}
{\cal C}_d ( {\bf X}, {\bf r} )\; = \; - \; \gamma^2_d \;\; {\rm  ln}  \frac{|{\bf X} - {\bf r}| }{R} \;.
\label{goal2}
\end{equation}
Combining  Eqs. (\ref{goal1}) and (\ref{goal2}), we obtain the desired sign prescription in Eq. (\ref{rfond}) 
and  conclude that the sign in Eq. \eqref{flowuredu} (or  the first equation in Eq. \eqref{rgdiff2d}), 
is a consequence of the dimensionality of the system.

\end{document}